\pdfoutput=1

\documentclass[fleqn,usenatbib]{mnras}

\usepackage{txfonts} 
\usepackage[T1]{fontenc}
\usepackage{amssymb}
\usepackage{graphicx}
\usepackage{multirow}
\usepackage[usenames, dvipsnames]{xcolor}
\usepackage{hyperref}
\usepackage[all]{hypcap}
\hypersetup{
    colorlinks,
    linkcolor={red!50!black},
    citecolor={blue!50!black},
    urlcolor={blue!80!black}
}

\let\la=\lesssim

\title[Decameter CRRLs towards Cas~A]{LOFAR observations of decameter carbon radio recombination lines towards Cassiopeia~A}

\author[P.~Salas et al.]{
P.~Salas$^{1}$\thanks{E-mail: psalas@strw.leidenuniv.nl}, 
J.~B.~R. Oonk$^{1,2}$, 
R.~J.~van~Weeren$^{3}$, 
F.~Salgado$^{1}$, 
L.~K.~Morabito$^{1}$,\newauthor
M.~C.~Toribio$^{1}$, 
K.~Emig$^{1}$, 
H.~J.~A.~R\"ottgering$^{1}$ and 
A.~G.~G.~M.~Tielens$^{1}$
\\
$^{1}$Leiden Observatory, Leiden University, P.O. Box 9513, NL-2300 RA Leiden, The Netherlands\\
$^{2}$Netherlands Institute for Radio Astronomy (ASTRON), Postbus 2, 7990 AA Dwingeloo, The Netherlands\\
$^{3}$Harvard-Smithsonian Center for Astrophysics, 60 Garden Street, Cambridge, MA 02138, USA
}

\date{Accepted XXX. Received YYY; in original form ZZZ}

\begin{document}
\label{firstpage}
\pagerange{\pageref{firstpage}--\pageref{lastpage}}
\maketitle

\begin{abstract}

We present a study of carbon radio recombination lines towards Cassiopeia~A using LOFAR observations in the frequency range $10$--$33$~MHz. 
Individual carbon $\alpha$ lines are detected in absorption against the continuum at frequencies as low as $16$~MHz. 
Stacking several C$\alpha$ lines we obtain detections in the $11$--$16$~MHz range.
These are the highest signal-to-noise measurements at these frequencies.
The peak optical depth of the C$\alpha$ lines changes considerably over the $11$--$33$~MHz range with the peak optical depth decreasing from $4\times10^{-3}$ 
at $33$~MHz to $2\times10^{-3}$ at $11$~MHz, while the line width increases from $20$~km~s$^{-1}$ to ${\sim150}$~km~s$^{-1}$.
The combined change in peak optical depth and line width results in a roughly constant integrated optical depth.
We interpret this as carbon atoms close to local thermodynamic equilibrium.

In this work we focus on how the $11$--$33$~MHz carbon radio recombination lines can be used to determine the gas physical conditions.
We find that the ratio of the carbon radio recombination lines to that of the $158$~$\mu$m [CII] fine-structure line is a good thermometer, while the ratio between low frequency carbon radio recombination lines provides a good barometer.
By combining the temperature and pressure constraints with those derived from the line width we are able to constrain the gas properties (electron temperature and density) and radiation field intensity. 
Given the $1\sigma$ uncertainties in our measurements these are; $T_{e}\approx68$--$98$~K, $n_{e}\approx0.02$--$0.035$~cm$^{-3}$ and $T_{r,100}\approx1500$--$1650$~K.
Despite challenging RFI and ionospheric conditions, our work demonstrates that observations of carbon radio recombination lines in the $10$--$33$~MHz range can provide insight into the gas conditions.

\end{abstract}


\section{Introduction}
\label{sec:intro}

Radio recombination lines (RRLs) are an important diagnostic tool to study the properties of the interstellar medium (ISM) in galaxies \citep[e.g.,][]{Gordon2009}. 
The population of carbon ions in a given $n$ level is determined by the gas density, temperature and radiation field, as well as the atomic physics involved \citep[e.g.,][]{Shaver1975,Watson1980,Salgado2016a}. 
By comparing the optical depth of carbon radio recombination lines (CRRLs) for a set of levels the gas properties can be determined. 

Low frequency CRRLs have been observed towards a number of galactic sources \citep{Konovalenko1984,Erickson1995,Roshi2000,Kantharia2001} and in particular 
against the bright radio source Cassiopeia~A \citep[Cas~A, 
e.g.][]{Konovalenko1981c,Ershov1982,Konovalenko1984,Anantharamaiah1985,Lekht1989,Payne1989,Anantharamaiah1994,Payne1994,Stepkin2007,Asgekar2013,Oonk2017}. 
Towards Cas~A three velocity components have been identified in CRRL emission and absorption \citep[e.g.,][]{Payne1989}, which correspond to gas located in the 
Perseus arm of the Galaxy and the Orion spur. 
Recent analysis by \citet{Oonk2017} has shown that the Perseus arm gas traced by low frequency CRRLs has a temperature of $\sim85$~K, an electron density of 
$\sim0.04$~cm$^{-3}$ and large column densities ($N_{\rm{H}}\sim10^{22}$~cm$^{-2}$). 
These properties suggest that the gas traced by CRRLs is in the interface between atomic \citep[e.g.,][]{Shuter1964,Davies1975,Bieging1991,Schwarz1997} and molecular gas \citep[e.g.,][]{Liszt1999,Mookerjea2006,Kilpatrick2014}. 
The data used by \citet{Oonk2017} consisted of observations between $300$--$390$~MHz and $33$--$78$~MHz obtained with the Westerbork radio telescope (WSRT) and the low frequency array (LOFAR) respectively.

Previous low frequency CRRL observations and models have suggested that at lower frequencies ($\nu<33$~MHz) the integrated optical depth of the CRRLs increases with increasing principal quantum number \citep[e.g.,][]{Payne1994}. This would make observations below $33$~MHz particularly interesting, since (i) given the large integrated optical depths involved the lines should be easily detected, and (ii) at these low frequencies, collisional and radiation broadening of RRLs provide insight into the physical conditions of the emitting gas \citep{Shaver1975,Salgado2016a,Salgado2016b}.

Low-frequency observations are hindered by ionospheric phase distortions, scintillation, and strong radio frequency interference (RFI). 
Below $20$~MHz these observations are very difficult due to the plasma frequency cutoff which is typically located around $10$~MHz, but varies depending on the ionospheric conditions \citep[e.g.,][]{Budden1985,Fields2013}. Nevertheless, CRRLs have been detected at frequencies as low as $12$~MHz \citep{Konovalenko2002b}.

Previous ${\nu<33}$~MHz studies have been carried out with the UTR-2 telescope in Ukraine \citep{Braude1978}. 
Using this telescope \citet{Konovalenko1980} reported the detection of a spectral line at $26$~MHz towards Cas~A, which was later identified as a CRRL with principal quantum number $n=631$ 
\citep{Blake1980,Konovalenko1981c}. 
They also reported the detection of six C$\alpha$ lines between $16.7$ and $29.9$~MHz \citep{Konovalenko1984}, and more recently showed a spectrum of five C$\alpha$ lines around $20$~MHz and a number of C$\beta$ lines \citep{Konovalenko2002a}.
These studies showed that the line peak optical depth decreases while the line width increases when the frequency decreases. 
\citet{Stepkin2007} reported the detection of C$\alpha$, C$\beta$, C$\gamma$ and C$\delta$ lines towards Cas A at $26$~MHz. 
Their detection of a C$\delta$ line sets the record for the largest bound atom ever detected with a principal quantum number $n\sim1000$.

LOFAR, operating at $10$--$240$~MHz \citep{vanHaarlem2013} provides a new opportunity to study RRLs at frequencies down to $10$~MHz. 
Due to its large bandwidth, hundreds of lines can be detected in a single observation. 
This opens up the possibility to study a broad range in principal quantum number with the same telescope. 
The LOFAR low band antenna (LBA) operates in the frequency range $10$--$90$~MHz. 
Previous LOFAR studies have focused on the higher frequency range ($33$--$70$~MHz) of the LBA \citep{Asgekar2013,Oonk2014,Morabito2014,Oonk2017} as the sensitivity of the LBA peaks in this frequency range \citep{vanHaarlem2013}. 

In this paper we report on LOFAR LBA observations of Cas~A between $10$ and $33$~MHz, the lowest frequency range LOFAR can reach. 
Our aim is to (i) determine if observations with LOFAR in this frequency range can yield CRRL detections, (ii) test if the integrated optical depth of the lines at high principal quantum number increases as suggested by previous observations, (iii) determine how low frequency CRRL observations can be used to constrain the physical conditions of the cold ISM.

The observations and data reduction are described in Sect.~\ref{sec:obs}. 
The results are presented in Sect.~\ref{sec:results} and then compared with models and previous results in \ref{sec:analysis}. 
This is followed by our conclusions in Sect.~\ref{sec:conclusions}.

\section{Observations \& data reduction}
\label{sec:obs}

\subsection{LOFAR observations}
\label{ssec:lofobs}

Cas~A was observed with the LOFAR LBA for two separate runs on October 20 and 21, 2012 for a total integration time of $20$~hr. 
This data were taken as part of the Lofar Cassiopeia~A Spectral Survey  (LCASS, PI, J.~B.~R.~Oonk). 
An overview of the observations is given in Table~1. 
During the time of the observations $34$ Dutch LOFAR stations were available. 
For the L69891 observation 26 stations recorded good data, while for L69893 this number was reduced to 21 due to ongoing upgrades. 
The entire $10$--$33$~MHz range was covered with $195.3125$~kHz wide subbands. 
Each subband had a total of $512$ channels, providing a channel width of $\sim380$~Hz or $\sim4$--$12$~km~s$^{-1}$.

\begin{table}
\begin{center}
\caption{LBA low observations}
\begin{tabular}{lllll}
\hline
\hline
\hline
Observations ID                 & L69891, L69893 \\
Integration time per visibility & 1~s \\
Observation dates               & 20, 21 October 2012 (15:00-01:00 UT) \\
Total on-source time            & 10 hr, 10 hr \\
Correlations                    & XX, XY, YX, YY \\
Frequency setup                 & 10--33~MHz full coverage \\
Bandwidth per subband           & 195.3125~kHz \\
Channels per subband            & 512 \\
Channel width                   & $4$--$12$~km~s$^{-1}$ \\
\hline
\hline
\end{tabular}
\label{tab:observations}
\end{center}
\end{table}

\begin{figure*}
\begin{center}
\includegraphics[angle=90,width=0.99\textwidth]{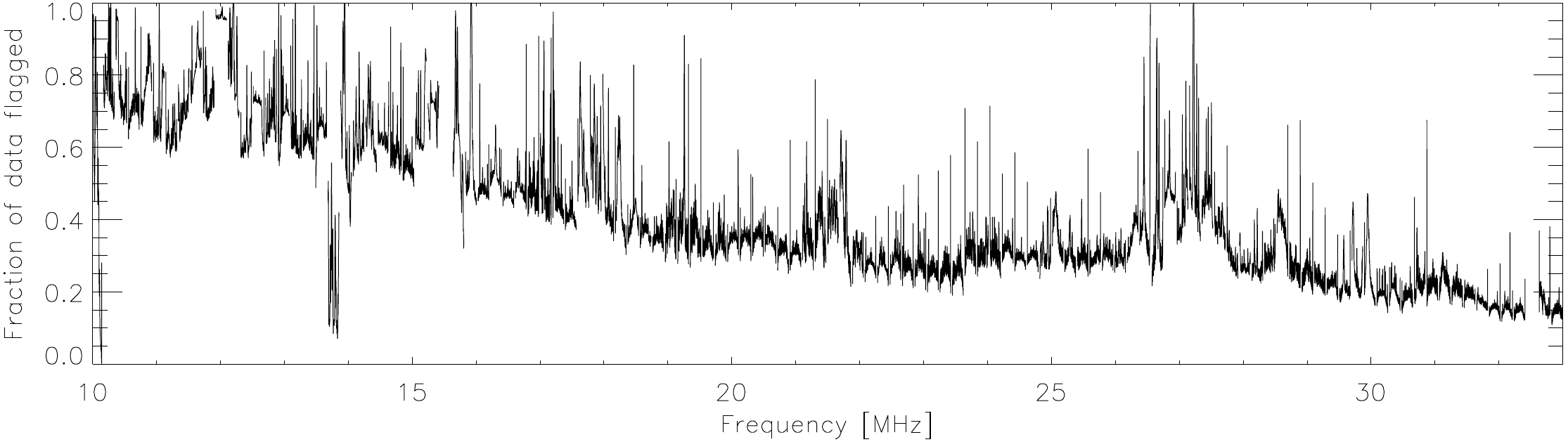}
\end{center}
\caption{Fraction of data flagged due to RFI or bad calibration solutions as function of frequency for the L69891 dataset (results for L69893 are very 
similar). The frequency width is the same as the one used in the observations, i.e., $380$~Hz.}
\label{fig:flags}
\end{figure*}

The first and last $25$ channels of each subband were flagged due to the bandpass roll-off. 
To mitigate the effect of RFI, we flagged the data using \emph{AOFlagger} \citep{Offringa2010,Offringa2012}. 
The percentage of data flagged due to RFI varied drastically across the band from about $20$\% at $30$~MHz to $40$\% at $20$~MHz and $70$\% at $15$~MHz, see Figure~\ref{fig:flags}. 
We also removed the last three hours of both datasets due to severe scintillation, although this part of the data was less affected by RFI by a factor of two to three.
We note that especially the first part of the observations were heavily affected by RFI, while the situation improved considerably after 23:00 UT for both datasets. 
Although the data were affected by scintillation after 22:00 UT, it suggests that night time observations are required to avoid severe RFI below $30$~MHz.

The data were then calibrated against a $11\farcs2\times9\farcs8$ resolution $69$~MHz model of Cas~A \citep{Oonk2017}. 
The calibration was performed with the New Default Processing Pipeline (\emph{NDPPP}) package which is part of the standard LOFAR software. 
Below $16$~MHz we could not obtain good gain solutions for several time ranges during the observations. 
These time ranges were also flagged and thus increase the percentage of flagged data (Figure~\ref{fig:flags}). 
Most likely, these periods correspond to severe ionospheric distortions. 
These are expected at low frequencies since the observations are carried out not far from the plasma cutoff frequency. 
However, signal from Cas~A was observed all the way down to $10$~MHz for certain time ranges.
The LOFAR bandpass is very smooth and, except for the bandpass roll-off regions (about 5 to 10 percent on either side of a subband) is well fit by a low order polynomial. 
To estimate how well a polynomial corrects the subband bandpass we also derived a bandpass calibration using Cygnus~A. 
We find that the results obtained from a polynomial are equivalent to those obtained with Cygnus~A. 
However, given that brightness of Cygnus~A is comparable to that Cas~A at these frequencies, using Cygnus~A leads to a slightly higher spectral noise (by about a factor $\sqrt2$). 
We therefore use a low order polynomial to correct our subband spectra for the bandpass shape.

Image cubes were made with \emph{AWImager} \citep{Tasse2013} by splitting and imaging one channel at a time. 
The images were convolved with a Gaussian beam with a size ranging between $23\arcmin$~at $10$~MHz and $5\arcmin$~at $30$~MHz.
Given the size of the convolving beam the background source is unresolved \citep[Cassiopeia~A has a radius of $\sim2.5\arcmin$ at $74$~MHz, e.g.,][]{DeLaney2014,Oonk2017}.
Spectra were then extracted from a tight box around the source at each channel.
We combined the spectra from the two different runs by weighting with the fraction of usable data at the corresponding frequency. 
Channels whose amplitude deviated more than $1\sigma$ from the median were removed. 
To determine the median we used a running median window filter with a box size of five channels.
$\sigma$ was determined using Tukey's biweight \citep[e.g. Equation (9) in][]{Beers1990}.
This typically resulted in less than $10\%$ of the data being flagged.
We also discard any channels where the percentage of flagged data deviated more than $5\%$ from the median amount of flagged data in the subband (see the bottom panel of Figure~\ref{fig:spectra}).

In each subband we estimate the continuum level by fitting a linear function to line free channels. 
Then the spectra are converted to optical depth units using \citep[e.g.,][]{Oonk2014}
\begin{equation}
 \tau_{\nu}=I_{\nu}/I_{\nu}^{\rm{cont}}-1.
 \label{eq:tau}
\end{equation}
Here $I_{\nu}$ is the spectrum extracted from the data cubes and $I_{\nu}^{\rm{cont}}$ is the continuum determined from line free channels.
Of the $286$ C$\alpha$ lines observable in the $10$--$33$~MHz range, $219$ were observed. 
Missing lines lie in the subband gaps or close to flagged edge channels.

\subsection{Herschel PACS archival data}

Cas~A was observed with the Photoconductor Array Camera and Spectrometer \citep[PACS, ][]{Poglitsch2010} instrument onboard the Herschel space observatory 
\citep{Pilbratt2010} during January $2011$ (observation IDs: $1342212243$--$1342212260$). These observations consist of nine footprints, each with $5\times5$ 
spaxels of $9\farcs4\times9\farcs4$. These nine PACS footprints cover $\sim20\%$ of Cas~A. The observations were made using range spectrography scans between 
$140$--$210$~$\mu$m with a $30$~km/s sampling near the $158$~$\mu$m line. To remove the instrument response, the chopping/nodding mode was used. 
However, the off source scans show significant line emission at the same velocity as the on source spectra. Because of this we used the on source spectra 
without removing the off source spectra. To separate on and off-source scans \emph{HIPE} version 14.0.0 was used \citep{Ott2010}.

\subsection{CRRL Stacking}
\label{ssec:stack}

\begin{figure}
 \begin{center}
  \includegraphics[width=0.49\textwidth]{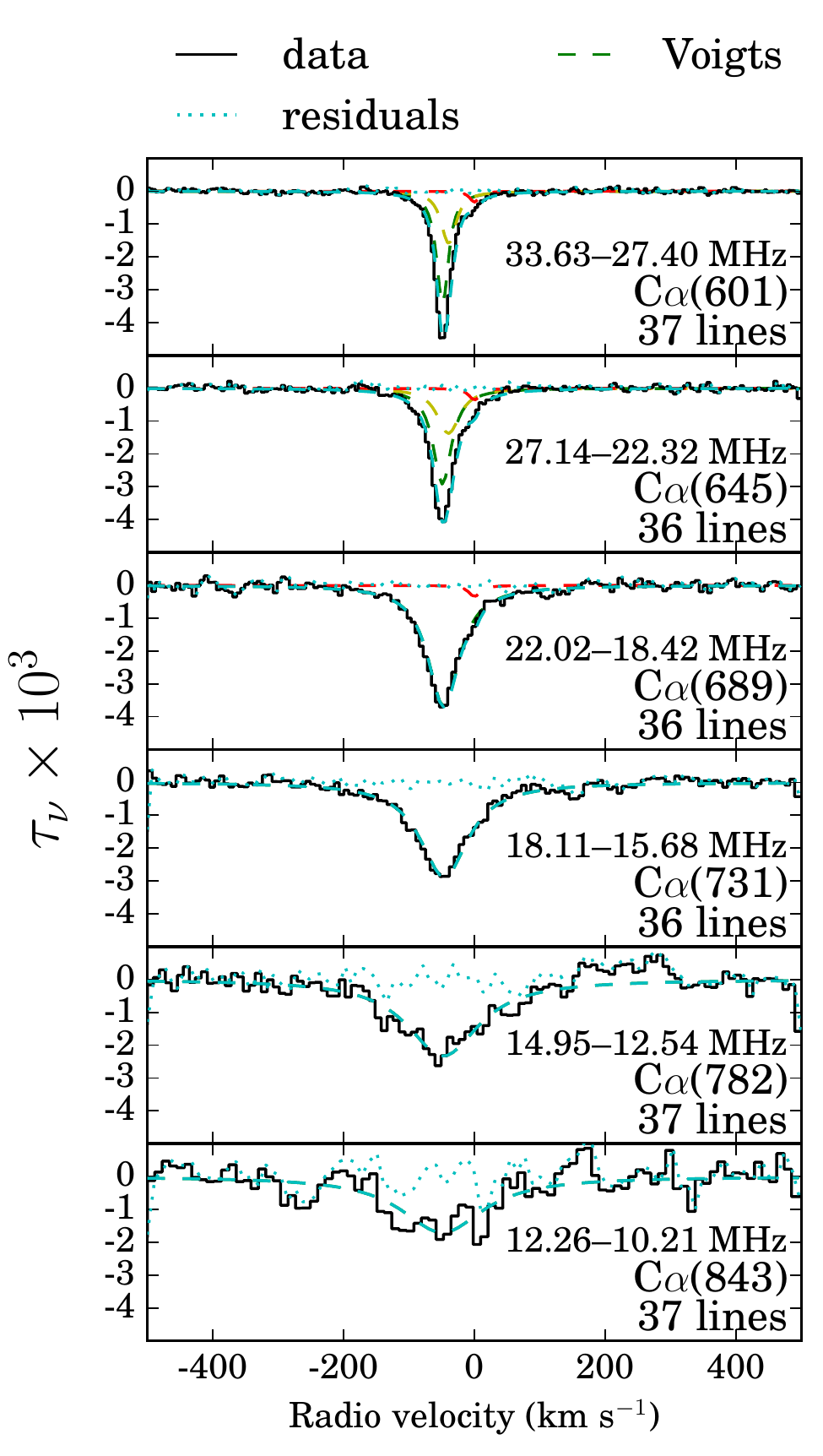}
 \end{center}
 \caption{Stacked C$\alpha$ spectra. 
 The {\it black steps} show the stacked spectra after processing, the {\it green}, {\it yellow} and {\it red dashed lines} show the best fitting Voigt profiles for the different components, the {\it cyan dashed line} shows the combined best fit Voigt profile and the {\it cyan dotted line} shows the residuals. 
 The spectra are shown in optical depth units.}
 \label{fig:calpha}
\end{figure}

Since the $10$--$33$~MHz range is heavily populated by CRRLs we used a procedure similar to that of \cite{Stepkin2007} to stack individual lines and obtain robust line profiles.
This stacking procedure also helps in removing residual bandpass structure.
First we searched the spectra for C$\alpha$ and C$\beta$ lines that were not blended with other lines, were not heavily affected by RFI and were far from the subband bandpass roll-off. 
We stacked these lines in the frequency ranges indicated in Figures~\ref{fig:calpha} and \ref{fig:cbeta}.
The stacking was performed by interpolating to a regular grid in velocity, with a bin size equal to the coarsest velocity resolution of the spectra included in the stack. 
Each individual spectrum was weighted by the inverse of its channel to channel variance.

\begin{figure}
 \begin{center}
  \includegraphics[width=0.49\textwidth]{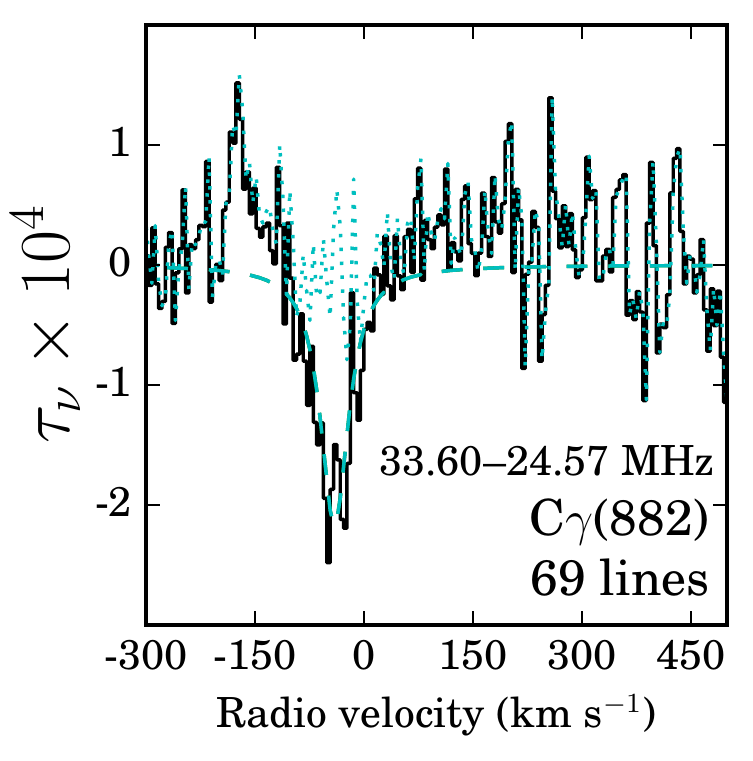}
 \end{center}
 \caption{Stacked C$\gamma$ spectrum. 
 The {\it black steps} show the stacked spectra after processing, the {\it cyan dashed line} shows the best fit Voigt profile and the {\it cyan dotted line} shows the residuals. 
 This is the same as in Figure~\ref{fig:calpha}.}
 \label{fig:cgamma}
\end{figure}

The stacked spectra were then fitted with a Voigt profile centred at $-47$~km~s$^{-1}$. 
Using the results of the fit to the stacked C$\alpha$ and C$\beta$ lines we then subtracted the best fit Voigt profile from each spectrum.
In the C$\alpha$ and C$\beta$ subtracted spectrum we searched for C$\gamma$ lines that were far from the roll-off of the subband bandpass, not blended with C$\delta$ lines and not heavily affected by RFI. 
These C$\gamma$ lines were stacked in $4$ frequency ranges. 
Of these ranges, only one yields a detection, see Figure~\ref{fig:cgamma}. 
In the remaining frequency ranges the signal-to-noise is below $3\sigma$. 
The C$\gamma$ stacks were fitted with a Voigt profile. 
The best fit C$\gamma$ Voigt profile was also removed from the C$\alpha$ and C$\beta$ free spectra. 
In the residual spectra we then searched for C$\delta$ lines by stacking all the available transitions in $2$ frequency ranges. 
These yielded non detections with $3\sigma$ upper limits on the peak optical depth of $10^{-4}$ for ${n=1020}$ and $2\times10^{-4}$ for ${n=1248}$. 
The C$\delta$ stack was also removed from the spectra. 
After this we performed a baseline correction on the line removed spectra using a polynomial of order $0$. 
Using the baseline corrected spectra we repeated the stacking of the lines. 
This was repeated $5$ times increasing the polynomial order by $1$ in each step. 
Finally, spectra with only one kind of transition were obtained by removing the corresponding best fit Voigt profiles from the spectra. 
The final C$\alpha$, C$\beta$ and C$\gamma$ spectra are shown in Figures~\ref{fig:calpha}, \ref{fig:cbeta} and \ref{fig:cgamma} respectively. 
In the final spectra, lines which were partially flagged due to RFI were also included.

\begin{figure*}
\begin{center}
\includegraphics[angle=90,trim=0cm 0cm 0cm 0cm,width=0.49\textwidth]{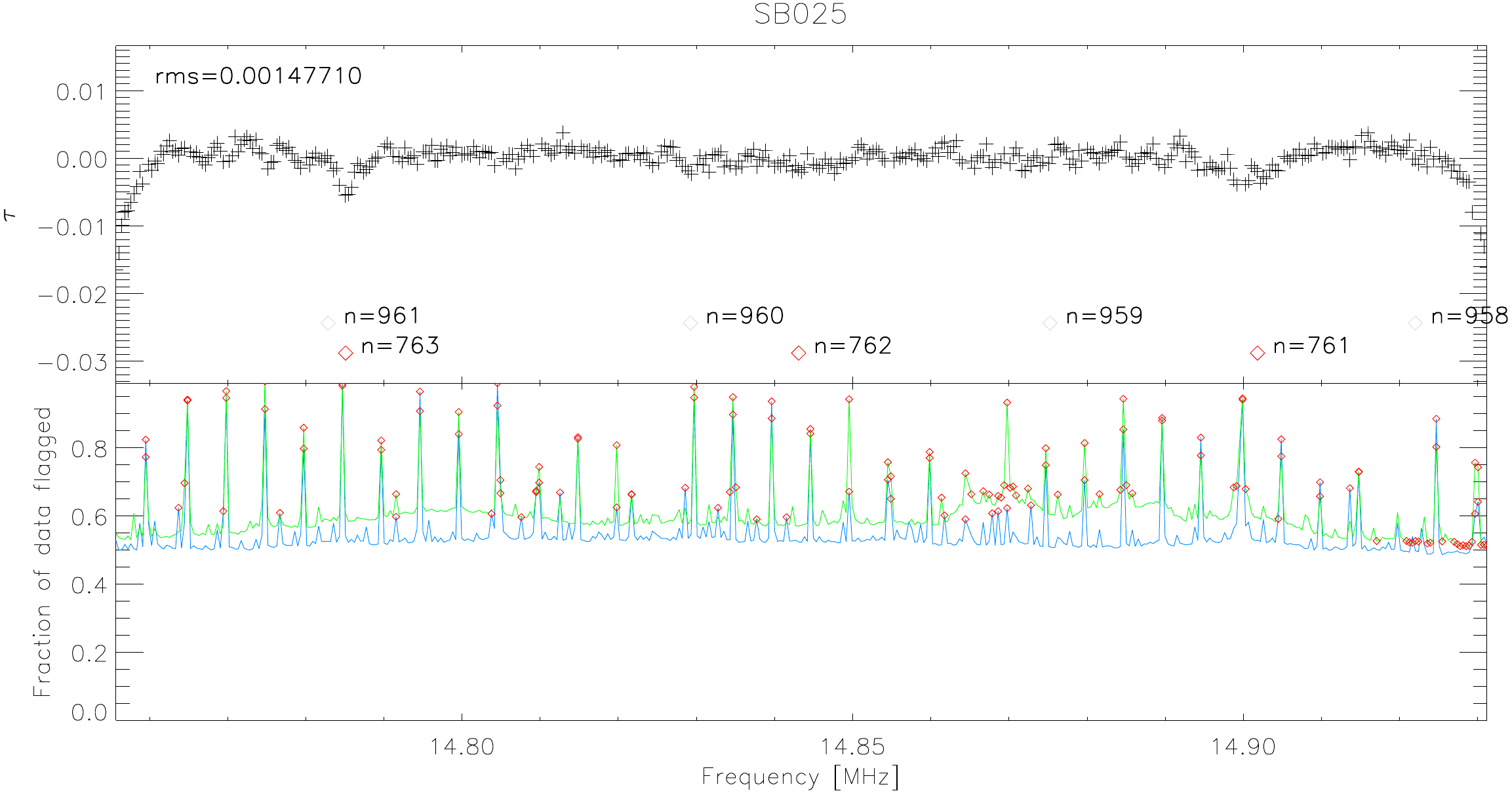}
\includegraphics[angle=90,trim=0cm 0cm 0cm 0cm,width=0.49\textwidth]{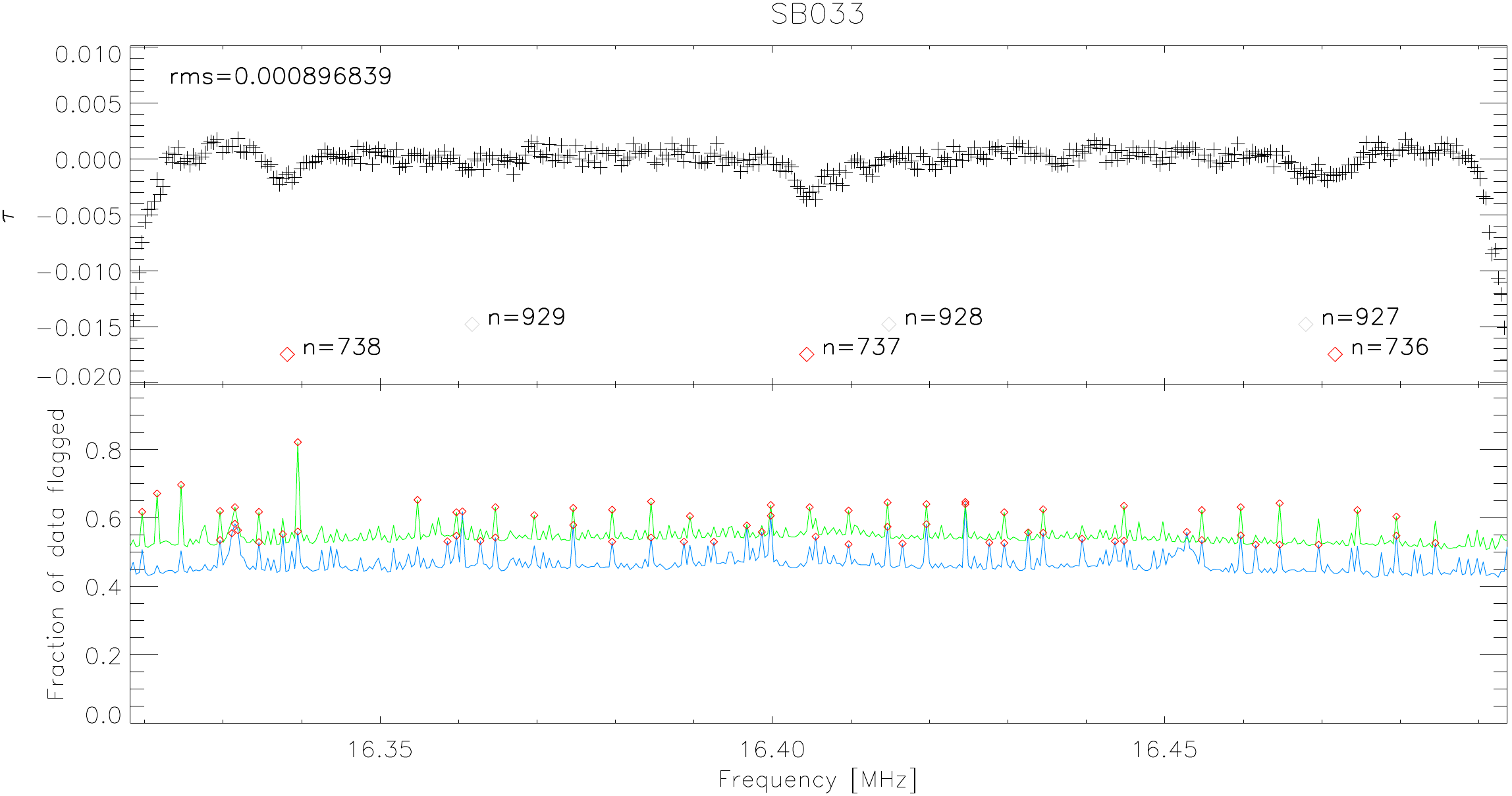}
\end{center}
\caption{Spectra of two individual subbands, $25$ and $33$. 
The {\it top} panels show the optical depth $\tau$ as function of frequency. 
The {\it red diamonds} show the location of C$\alpha$ lines, while the empty labels those of C$\beta$. 
The {\it bottom} panels show the percentage of data flagged for the L69891 and L69893 datasets in blue and green respectively. 
{\it Red diamonds} mark channels which were removed because their flagging percentage is larger than $5\%$ that of the subband mean.}
\label{fig:spectra}
\end{figure*}

\citet{Payne1994} showed that a finite bandwidth combined with a baseline removal process can cause systematic biases in the observed line profiles.
To test whether the stacking and baseline removal procedure combined with the correlator setup used introduces any biases in the measured line properties we performed a test using synthesised spectra. 
The synthetic spectra are generated with known line properties. 
These synthetic spectra are stacked in the same way as the observed data, including the baseline and line removal (e.g., removing the C$\beta$, C$\gamma$ and C$\delta$ lines from the C$\alpha$ stack) steps.
Then, by comparing the known input model properties with the measured line properties after stacking we can determine if the measured line properties are distorted by the stacking procedure. 
The tests show that without baseline and line removal the line properties are up to $40\%$ different from those of the input model. 
If we apply the baseline and line removal during stacking the difference is less than $15\%$, where the highest $n$ data is the hardest to recover. 
From this we conclude that baseline corrections on scales larger than the line width and line removal help recover the line profiles.
These tests also show that we can recover the line properties of the dominant velocity component at $-47$~km~s$^{-1}$. 
To do so we need to fit three velocity components up to $n\approx650$ and two components from $n=650$ to $700$.
The details of these simulations can be found in the Appendix.

\section{Results}
\label{sec:results}

We display examples of individual subband spectra in Figure~\ref{fig:spectra}. 
Individual C$\alpha$ lines are visible above $16$~MHz. 
The peak optical depth ranges from $4.6\times10^{-3}$ at $30$~MHz to $2\times10^{-3}$ at $11$~MHz. 
Individual C$\beta$ lines are visible above ${\nu\gtrsim28}$~MHz.

\begin{figure}
 \begin{center}
  \includegraphics[width=0.49\textwidth]{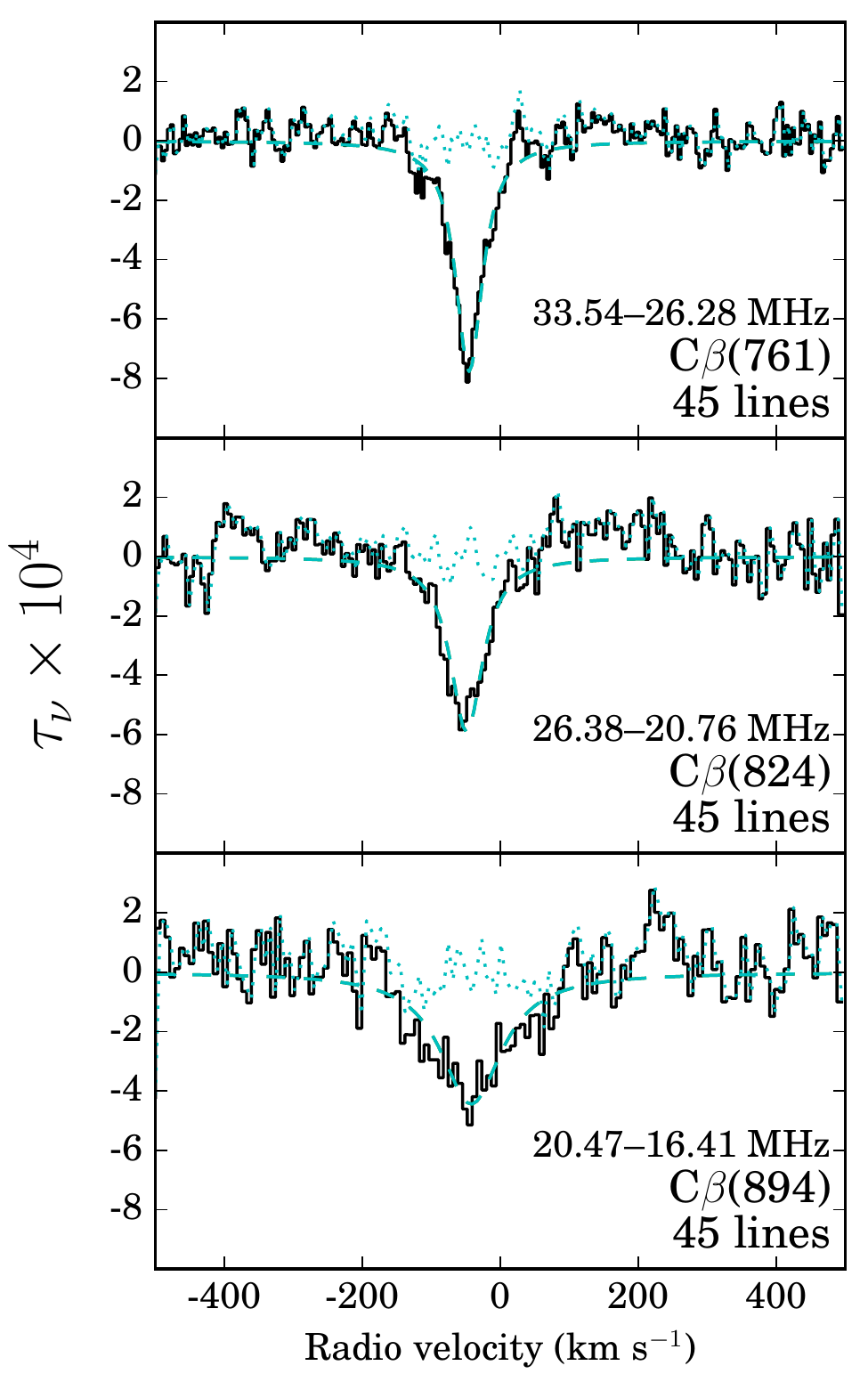}
 \end{center}
 \caption{Stacked C$\beta$ spectra.
 The {\it black steps} show the stacked spectra after processing, the {\it cyan dashed line} shows the best fit Voigt profile and the {\it cyan dotted line} shows the residuals.
 This is the same as in Figure~\ref{fig:calpha}.}
 \label{fig:cbeta}
\end{figure}

The line profiles obtained after stacking are shown in Figures~\ref{fig:calpha}, \ref{fig:cbeta} and \ref{fig:cgamma}. 
These Figures show how the line width increases while the peak optical depth decreases towards lower frequencies. 
The C$\alpha$ stacks show detections down to the lowest frequency of $10.96$~MHz, with a signal-to-noise ratio of $3.7\sigma$. 
To our knowledge this is the lowest frequency detection of a C$\alpha$ line to date.

The line centroid is found to be offset by $-47$~km~s$^{-1}$ with respect to the local standard of rest, $v_{\rm{LSR}}$, as expected from all previous CRRL 
observations in the direction of Cas~A. This velocity is coincident with the strongest HI absorption component observed against Cas~A and it is associated with 
gas in the Perseus arm of our Galaxy \citep[e.g.,][]{Davies1975}. Spectra of CRRLs at higher frequencies, where the line broadening is not as severe and the 
velocity resolution is better, show that along this line of sight $2$ additional velocity components can be identified. One is at $-38$~km~s$^{-1}$, also 
associated with the Perseus arm of the Galaxy, and the other is at $0$~km~s$^{-1}$, associated with the Orion spur \citep[e.g.,][]{Payne1989}.

\subsection{Line fitting}

\subsubsection{CRRLs}
\label{sssec:crrls}

Based on the results of the fits to the simulated line profiles we use a different number of Voigt profiles to fit the stacked lines. 
The number of profiles is determined by requiring that the residuals are reduced by more than a factor of two while still recovering reasonable line parameters. 
For ${n<650}$ using three profiles, two for the Perseus arm and one for the Orion spur components, results in the lowest residuals. 
The Doppler widths of the $-47$, $-38$ and $0$~km~s$^{-1}$ components are fixed at $3.4$, $6.8$ and $2.5$~km~s$^{-1}$ respectively \citep{Oonk2017}. 
For ${650<n<700}$, we fit two Voigt profiles, one for the Perseus component and one for the Orion component. 
For ${n>700}$ a single component is used to fit the line profile. 
The best fit parameters are given in Table~\ref{tab:vpars}. 
Here we see that even when we fit a single component the observed C$\alpha$ line centre is close to $-47$~km~s$^{-1}$. 
This indicates that this velocity component still dominates at the lower frequencies. 
While the observations clearly reveal the presence of CRRLs up to $n=843$ (see Figure~\ref{fig:calpha}), corresponding to $\nu=10$--$12$~MHz, quantitative analysis of this data is limited by recovery of the line wings (see 
Appendix~\ref{app:sim}). 
We have ignored the C$\alpha(843)$ data in further analysis.

To test the fit results we varied the noise properties of the stacked spectra and repeated the fit. 
To do this we subtract the best fit Voigt profiles from the stacked spectra. 
The best fit Voigt profiles are then added to a spectra with a different noise level. 
The noise level was randomly drawn from a Gaussian distribution with a standard deviation equal to the rms in the stack. 
The stacks with the different noise levels were then fit again. 
We repeated this $1000$ times for every stacked spectrum. 
The measured line properties varied little between different realisations of the noise. 
For each stack we made histograms of the measured line properties. 
The distribution of the line parameters showed Gaussian distributions for the line width and integrated optical depth. 
The standard deviation of these distributions are comparable to the errors given for the line parameters in Table~\ref{tab:vpars}.

\begin{table*}
\begin{center}
\caption{Line parameters}
\begin{tabular}{cccccccc}
\hline
\hline
\hline
$n$ & $n$ range & Frequency & $I_{l}/I_{c}$   & Line centre   & Lorentz line width & Integrated optical depth & Spectral rms         \\
    &           & (MHz)     & $\times10^{-3}$ & (km~s$^{-1}$) & (km~s$^{-1}$)      & (Hz)                     & $\times10^{-4}$      \\
\hline
\multicolumn{8}{c}{C$\alpha$}                                                                                          \\
\hline                                                                                                                 
\multirow{3}{*}{601} & \multirow{3}{*}{580--621} & \multirow{3}{*}{30.23} & 3.44$\pm$0.05 & -48$\pm$3       & 17$\pm$2   & 11.9$\pm$0.1 & 0.4 \\
                     &                           &                        & 1.59$\pm$0.04 & -39$\pm$3       & 22$\pm$2   & 6.8$\pm$0.1  & 0.4 \\
                     &                           &                        & 0.33$\pm$0.06 & 0$\pm$3         & 13$\pm$3   & 0.9$\pm$0.1  & 0.4 \\
\multirow{3}{*}{645} & \multirow{3}{*}{623--665} & \multirow{3}{*}{24.46} & 2.92$\pm$0.07 & -49$\pm$4       & 29$\pm$2   & 12.7$\pm$0.3 & 0.5 \\
                     &                           &                        & 1.38$\pm$0.06 & -39$\pm$4       & 35$\pm$2   & 7.3$\pm$0.3  & 0.5 \\
                     &                           &                        & 0.3$\pm$0.1   & 0$\pm$4         & 23$\pm$4   & 1.1$\pm$0.2  & 0.5 \\
\multirow{2}{*}{689} & \multirow{2}{*}{668--709} & \multirow{2}{*}{20.07} & 3.6$\pm$0.1   & -46$\pm$5       & 58$\pm$1   & 22.8$\pm$0.4 & 0.8 \\
                     &                           &                        & 0.3$\pm$0.1   & 0$\pm$5         & 20$\pm$8   & 0.7$\pm$0.4  & 0.8 \\
731                  & 713--748                  & 16.80                  & 2.88$\pm$0.07 & -46$\pm$6       & 89$\pm$3   & 22.7$\pm$0.5 & 1.4 \\
782                  & 760--806                  & 13.73                  & 2.3$\pm$0.1   & -47$\pm$8       & 132$\pm$12 & 22$\pm$1     & 3.0 \\
843                  & 812--863                  & 10.96                  & 1.7$\pm$0.1   & -48$\pm$10      & 153$\pm$26 & 15$\pm$1     & 4.0 \\
\hline                                                                                                                 
\multicolumn{8}{c}{C$\beta$}                                                                                           \\
\hline                                                                                                                 
761                 & 731--793                   & 29.74     & -0.78$\pm$0.02 & -45$\pm$4     & 45$\pm$2           & 5.6$\pm$0.1         & 0.5 \\
824                 & 792--858                   & 23.43     & -0.58$\pm$0.04 & -48$\pm$5     & 54$\pm$5           & 3.9$\pm$0.2         & 0.8 \\              
894                 & 862--928                   & 18.35     & -0.44$\pm$0.04 & -41$\pm$6     & 103$\pm$14         & 4.4$\pm$0.4         & 1.0 \\
\hline                                                                                                
\multicolumn{8}{c}{C$\gamma$}                                                                                          \\
\hline                                                                                                
882 & 836--928   & 28.62     & 0.21$\pm$0.01  & -39$\pm$4     & 44$\pm$7           & 1.5$\pm$0.1             & 0.5                   \\
\hline
\hline
\end{tabular}
\label{tab:vpars}
\end{center}
\end{table*}

\subsubsection{$158$~$\mu$m \textnormal{[CII]} line}

\begin{figure}
 \begin{center}
  \includegraphics[width=0.49\textwidth]{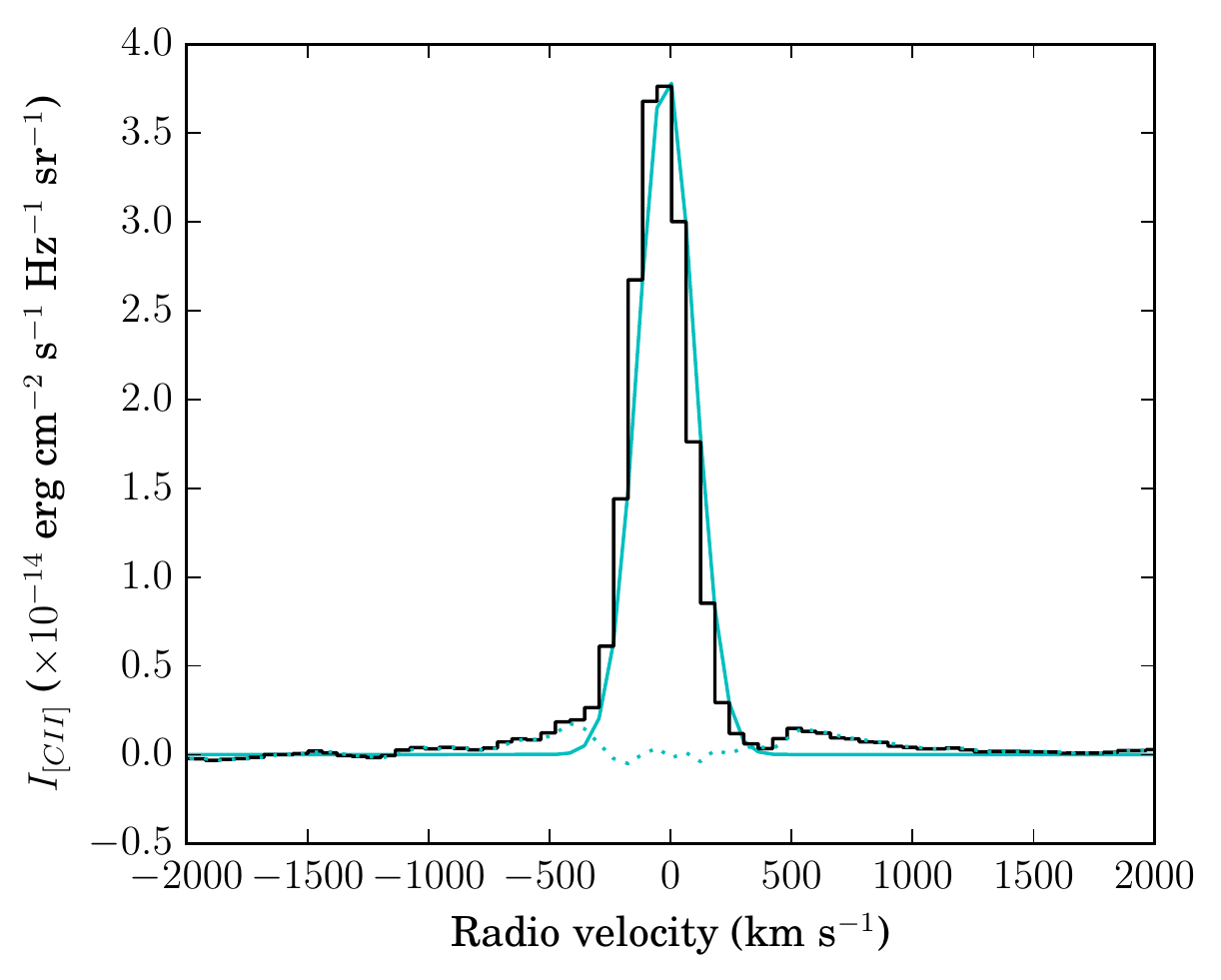}
 \end{center}
 \caption{[CII] $158$~$\mu$m line spectra obtained with PACS. 
 The line was averaged over the face of Cas~A after removing the continuum. 
 The {\it black steps} show the data, the {\it cyan solid line} shows the best fit Gaussian profile and the {\it dotted cyan line} shows the fit residuals.
 This is similar to Figure~\ref{fig:calpha}.}
 \label{fig:pacs}
\end{figure}

From each of the nine PACS footprints we extracted an average spectrum. 
We fitted the [CII] $158$~$\mu$m line in this average spectrum with a Gaussian.
To account for the continuum emission we fit a linear function to line free channels.
The source averaged spectra with the continuum subtracted is shown in Figure~\ref{fig:pacs}.
The line centroid is $-16\pm19$~km~s$^{-1}$, its full width at half maximum $218\pm3$~km~s$^{-1}$ and its amplitude $0.096\pm0.003$~Jy~arcsec$^{-2}$.
The average line intensity integrated over frequency over the face of Cas~A is $(7.0\pm0.2)\times10^{-5}$~erg~cm$^{-2}$~s$^{-1}$~sr$^{-1}$. 
Because the PACS footprints cover only $\sim20\%$ of the supernova remnant the uncertainty on the line intensity can be larger. 
Given the velocity resolution of the PACS observations the line is not resolved. 
This implies that the Perseus, Orion and any other velocity components present in [CII] appear as a single component. 
Even background [CII] emission could contaminate our spectra.
Observations of other tracers, with higher spectral resolution, such as the $21$~cm spin-flip transition of HI \citep{Bieging1991} and CO \citep{Kilpatrick2014}, show that the most prominent component is at $-47$~km~s$^{-1}$. 
Since we do expect a correlation between [CII], CO and HI \citep[e.g.,][]{Pineda2013}, the $158$~$\mu$m line intensity should be dominated by the contribution from the $-47$~km~s$^{-1}$ component.

\subsection{CRRL observed properties}

In the frequency range $10$--$33$~MHz two properties of the observed CRRL profiles can be used to determine the gas physical conditions, the line width and its 
integrated optical depth. Here we present the change in line properties, which will be later used to determine the gas physical conditions.

The line widths determined from fitting Voigt profiles to the CRRL stacks are shown in Figure~\ref{fig:lw}. 
Here we see how the lines get broader as the principal quantum number increases. 
The increase in line width causes the lines to become blended as frequency decreases. 
In the top panels of Figure~\ref{fig:calpha} it is possible to see the asymmetry in the line profile towards positive velocities due to the presence of additional gas. 
This is washed away as the dominant component gets broader. 
Our data in combination with the LCASS data \citep{Oonk2017} at higher frequencies clearly reveals the transition from Doppler dominated to Lorentz dominated line widths in the range $n=500$--$600$.

The integrated optical depths are shown in Figure~\ref{fig:itau}. 
Since for part of the low frequency spectra presented here ($n>700$) it is not possible to find a unique solution to separate the different velocity components we show the sum of the optical depths from the Perseus arm gas at $-47$ and $-38$~km~s$^{-1}$. 
This is done by adding the two components together when more than two Voigt profiles are fit, or by subtracting a $5\%$ contribution, corresponding to the Orion component, when one Voigt profile is fit. 
The LCASS points at $n>600$ show a fairly constant integrated optical depth as $n$ increases up to $800$.
In Figure~\ref{fig:itau} we also show the LCASS data \citep{Oonk2017} at higher frequencies, the latest compilation of integrated optical depths \citep{Kantharia1998} as well as the more recent C$\alpha$ data from \citet{Stepkin2007} and \citet{Sorochenko2010}.
The line width and integrated optical depth are used in the following section to constrain the gas properties.

\section{Analysis}
\label{sec:analysis}

In this section we analyse CRRLs below $33$~MHz in combination with the $158$~$\mu$m [CII] line. 
For the analysis we exclude the C$\alpha(843)$ line for the reasons given in Sect.~\ref{sssec:crrls}.

\subsection{Line width}
\label{ssec:lw}

\begin{figure}
 \begin{center}
  \includegraphics[width=0.49\textwidth]{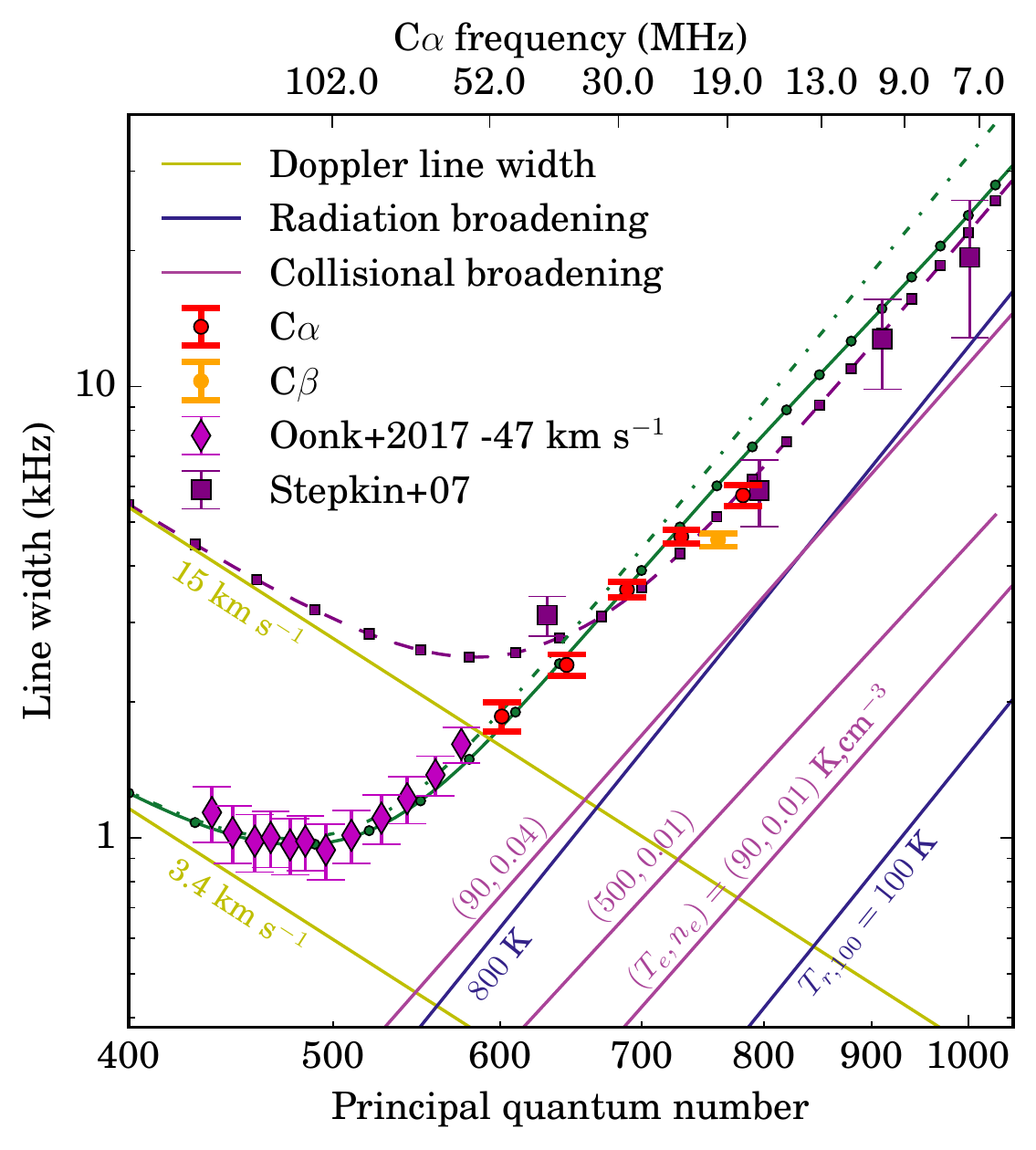}
 \end{center}
 \caption{Line width for the $-47$~km~s$^{-1}$ velocity component as a function of principal quantum number. 
 The {\it red} and {\it orange points} show the measured line widths from this work (Table~\ref{tab:vpars}). 
 The {\it purple diamonds} the line widths from \citet{Oonk2017} and the {\it purple squares} the C$\alpha$, C$\beta$, C$\gamma$ and C$\delta$ data from \citet{Stepkin2007}. 
 The {\it coloured solid lines} show the contribution from Doppler broadening ({\it yellow lines}), pressure broadening ({\it purple lines}) and radiation broadening ({\it blue lines}). 
 The {\it purple line with squares} shows the line width produced by the \citet{Stepkin2007} model.
 The {\it green dot dashed line shows the line width for a model with a Doppler line width of $3.4$~km~s$^{-1}$, $T_{e}=85$~K, $n_{e}=0.04$~cm$^{-3}$, and a power law radiation field with $\alpha=-2.6$ and $T_{r,100}=1400$~K \citep{Oonk2017}.}
 The {\it green line with circles } shows the line width for a model with the same physical conditions as the green dot dashed line, but in this case the radiation field is a combination of a power law with $T_{r,100}=800$~K and $\alpha=-2.6$ plus a contribution to the radiation field from Cas~A. 
 To model the contribution from Cas~A we use its observed flux density \citep{Vinyaikin2014}.
 We can see that the line width due to the radiation field from the observed flux density of Cas~A \citep{Vinyaikin2014} decreases faster than that of a power law with $\alpha=-1$ \citep[][line with squares versus line with circles]{Stepkin2007}.}
 \label{fig:lw}
\end{figure}

Two effects dominate the broadening of low frequency CRRLs. These are, pressure broadening, which depends on the electron temperature and density, and radiation 
broadening, which depends on the radiation field in which the atoms are immersed \citep{Shaver1975,Salgado2016a}. 
Both effects affect the Lorentzian wings of the line profile.
As mentioned in Sect.~\ref{ssec:stack} the line width at the low frequencies, where the lines are heavily blended, will be dominated by the most prominent component. 
In this case this is the $-47$~km~s$^{-1}$ component.
For principal quantum numbers below $\sim700$ the line profiles of the $-47$ and $-38$~km~s$^{-1}$ components can be decomposed.
For $n>700$, the line profile shows no deviations from a single Voigt, but at this point the $-47$~km~s$^{-1}$ is so broad that it dominates the line width.
In Figure~\ref{fig:lw} we present the line width for the $-47$~km~s$^{-1}$ velocity component as a function of principal quantum number over the full LBA frequency range. 
The higher frequency data for the $-47$~km~s$^{-1}$ component is taken from \citep{Oonk2017}. 
In Figure~\ref{fig:lw} we see that the line width increases with principal quantum number.

The coloured curves in Figure~\ref{fig:lw} show the contribution from pressure and radiation broadening on the lines. 
The line width due to pressure broadening is \citep[e.g.,][]{Salgado2016b},
\begin{equation}
 \Delta\nu_{p}=\frac{1}{\pi}10^{a}n_{e}n^{\gamma_{c}}\;\mbox{Hz}.
\end{equation}
Here, $n_{e}$ is the electron density, $n$ the principal quantum number, $a$ and $\gamma_{c}$ are constants that depend on the electron temperature of the gas, 
$T_{e}$. The values of $a$ and $\gamma_{c}$ are tabulated in \citet{Salgado2016b}. 
Using this model the maximum electron density of the gas is $0.3$~cm$^{-3}$ for the lowest temperature in our grid of models, i.e. $T_{e}=10$~K. 
The maximum allowed electron density decreases for higher temperatures.

Radiation broadening produces a line width given by \citep{Salgado2016b},
\begin{equation}
 \Delta\nu_{r}=\frac{1}{\pi}\sum_{n\neq n^{\prime}}B_{n^{\prime}n}I_{\nu}\;\mbox{Hz}.
 \label{eq:lwtr}
\end{equation}
Here, $B_{n^{\prime}n}$ is the Einstein coefficient for stimulated emission and $I_{\nu}$ the intensity of the radiation field. At low frequencies 
the background radiation field is generally due to synchrotron emission and can be described by a power law $T_{r}\propto\nu^{\alpha}$. 
Using the available low frequency surveys \citet{Zheng2016} find that the spectral index of synchrotron emission below $408$~MHz is 
$\alpha_{\rm{MW}}=-2.52\pm0.02$.
Studies of low frequency CRRLs in the Milky Way usually assume that $\alpha=-2.6$ \citep[e.g.,][]{Payne1989,Kantharia1998,Oonk2017}. The difference in 
$\Delta\nu_{r}$ between using a power law index of $-2.52$ and $-2.6$ is less than $20\%$ for $n=400$--$1000$. 
Here we will adopt the value of $\alpha_{\rm{MW}}=-2.6$ to be consistent with previous work. 
In this case the line width due to radiation broadening becomes
\begin{equation}
 \Delta\nu_{r}\approx0.8\left(\frac{T_{r,100}}{1000\;\rm{K}}\right)\left(\frac{n}{600}\right)^{5.8}\;\mbox{kHz}.
 \label{eq:mwdnurad}
\end{equation}
Where, $T_{r,100}$ is the brightness temperature of the radiation field at $100$~MHz in units of K. 

Using Eq.~\ref{eq:mwdnurad} and the line widths for $n<750$ (see Sect.~\ref{sssec:crrls}) we constrain $T_{r,100}$ to $<2000$~K. 
This is a strict upper limit, since pressure broadening is not considered. 
An approximate lower limit can be obtained from the lowest brightness temperature of the Milky Way in a region of similar galactic latitude to Cas~A. 
Using the global sky model of \citet{Oliveira-Costa2008} we have a value of $T_{r,100}\gtrsim800$~K. 
However, the exact value is highly uncertain.
Fitting a power law to the change in line width with principal quantum number in the range $n=601$--$731$ we find that we can not discriminate between collisional or radiation broadening. 
This because both effects have a similar dependence on $n$.
\citet{Oonk2017} find a gas electron density of $0.04$~cm$^{-3}$, a temperature of $85$~K and a radiation field intensity at $100$~MHz of $1350$~K, for the $-47$~km~s$^{-1}$ velocity component.
If we use these results for the gas conditions then the relative contributions from pressure and radiation broadening are similar. 
This suggests that both pressure and radiation broadening are of importance in setting the behaviour of the line width along this line of sight.

If we combine the Doppler, pressure and radiation induced broadening terms and fit it to the data with frequencies greater than $30$~MHz, then we arrive at the following relation for $T_{e}$, $n_{e}$ and $T_{r,100}$;
\begin{equation} 
\left(\frac{n_{e}}{\;\rm{cm}^{-3}}\right)=\left(0.8-0.38\left(\frac{T_{r,100}}{1000\;\rm{K}}\right)\right)\left(\frac{T_{e}}{\rm{K}}\right)^{-0.46}.
\label{eq:lwrel}
\end{equation}
This expression is valid for this line of sight when $T_{r,100}\lesssim1600$~K and accurate to within $10\%$.
In Sect.~\ref{ssec:pat} we adopt this relation to constrain the gas properties.
In Sect.~\ref{ssec:pat} we also assume that this relation is valid for both the $-47$ and $-38$~km~s$^{-1}$ velocity components. 
This assumption is based on the similitude between the derived gas physical properties from both velocity components \citep{Oonk2017}.
The physical conditions in this case are derived using higher frequency data, where the line profile can be more reliably decomposed.

\citet{Stepkin2007} showed that using a power law radiation field with $T_{r,100}=3200$~K overestimated the line widths for $n>600$.
To solve this issue they argued that the contribution from Cas~A to the radiation field has to be taken into account when modelling the line widths. 
The spectrum of Cas~A has a turnover point close to $30$~MHz \citep[e.g.,][]{Braude1978,Vinyaikin2014}, which causes its flux density to decrease for frequencies lower than the turnover frequency.
This solved the apparent discrepancy between the observed line widths and the model predictions.
Here we make use of the broadening expressions derived by \citet{Salgado2016b}. 
These result in a $\sim30\%$ lower Lorentz line width with respect to the \citet{Shaver1975} expressions used by \citet{Stepkin2007}.
For $n\lesssim650$, the use of expressions which predict a smaller line width combined with our attempt to fit three velocity components results in line widths which are consistent at the $3\sigma$ level with a power law line broadening.
However, as our simulations show, recovering the correct line width for $n\la700$ depends on the number of components fit to the blended line profiles, which has no unique solution.
Additionally, for $n\geq800$ the line profiles are likely to be underestimated due to the severe line broadening. 
At some point the width of the lines within a single subband is such that it is no longer possible to find line-free channels.
We believe that the combination of these effects could mimic a deviation from a power law line broadening (e.g., a broken power law).

In Figure~\ref{fig:lw} we also show the line width when the contribution from Cas~A is taken into account. 
To include the contribution from Cas~A to the radiation field ($I_{\nu}$ in Eq.~\ref{eq:lwtr}) we add it to the radiation field from the Milky Way.
To model the intensity of the radiation field due to Cas~A at frequencies lower than $100$~MHz we use two different models.
One is the model by \citet{Stepkin2007}, in which a broken power law is used. 
This broken power law has a turnover at $\nu=26$~MHz, for frequencies below this point the spectral index is $\alpha=-1$.
Above the turnover the spectral index is the same as that from the Milky Way radiation field, i.e. $\alpha=-2.6$.
In this model the contribution from Cas~A to the radiation field at the cloud is $T_{r,100}\approx360$~K.
The second model we use is based on the observed flux density from Cas~A \citep{Vinyaikin2014}.
In this case the contribution from Cas~A is $T_{r,100}\approx360$~K, which is determined from a fit to the observed line widths.

\subsection{Integrated optical depth}
\label{ssec:itau}

The observed change in integrated optical depth with principal quantum number, $n$, can be used to constrain the properties of the CRRL emitting gas 
\citep[e.g.,][]{Dupree1971,Shaver1975,Walmsley1982,Ershov1984,Payne1989,Ponomarev1992,Kantharia1998,Salgado2016a,Salgado2016b,Oonk2017}. 
Here we compare the observed integrated optical depth to those predicted by the updated models of \citet{Salgado2016a,Salgado2016b}. 
In these models the level populations are fully determined by the atomic physics involved, including the effect of dielectronic capture \citep{Watson1980}. 
The level populations are obtained by self consistently solving the statistical equilibrium equations. 
Deviations from LTE in the level population are characterised by the departure coefficient $b_{n}$, while the contribution from stimulated emission to the line intensity is characterised by $\beta_{n^{\prime}n}$ \citep{Shaver1975,Salgado2016a}.

To model the line optical depth we assume that the absorbing gas is a plane parallel slab which completely covers the face of Cas~A.
We assume a constant electron temperature and density.
The assumption of constant temperature and density is justified since we are studying a small frequency range ($10$--$33$~MHz) and density filter effects should be small.
The assumption of a unity beam filling factor seems reasonable given the spatially resolved observations of \citet{Anantharamaiah1994}, which show emission all over the face of Cas~A \citep{Oonk2017}.
When we compare the integrated optical depth to the models, we use the combined optical depth of the $-47$ and $-38$~km~s$^{-1}$ velocity components.
We do this based on the detailed analysis at higher frequencies \citep{Payne1994,Oonk2017}, which suggests that both velocity components trace gas with similar temperature and density.
In this case, the combined integrated optical depth of both velocity components will show the same behaviour with $n$ scaled by the total emission measure.
These assumptions will be used in the rest of the analysis.

When solving the statistical level population problem using the models of \citet{Salgado2016a,Salgado2016b} we use atomic hydrogen and electrons as collisional partners and we chose to ignore collisions with molecular hydrogen since we expect their importance to be relatively small given the gas physical conditions \citep{Oonk2017}. 
We evaluate the models in a grid in $n_{e}$--$T_{e}$--$T_{r,100}$ space. 
For the electron density the grid spans the range $n_{e}=0.01$--$1.1$~cm$^{-3}$ in steps of $0.005$~cm$^{-3}$. 
The electron temperature is evaluated between $10$--$150$~K in steps of $5$~K. 
$T_{r,100}$ is evaluated at $800$, $1200$, $1400$, $1600$ and $2000$~K.

\begin{figure}
 \begin{center}
  \includegraphics[width=0.49\textwidth]{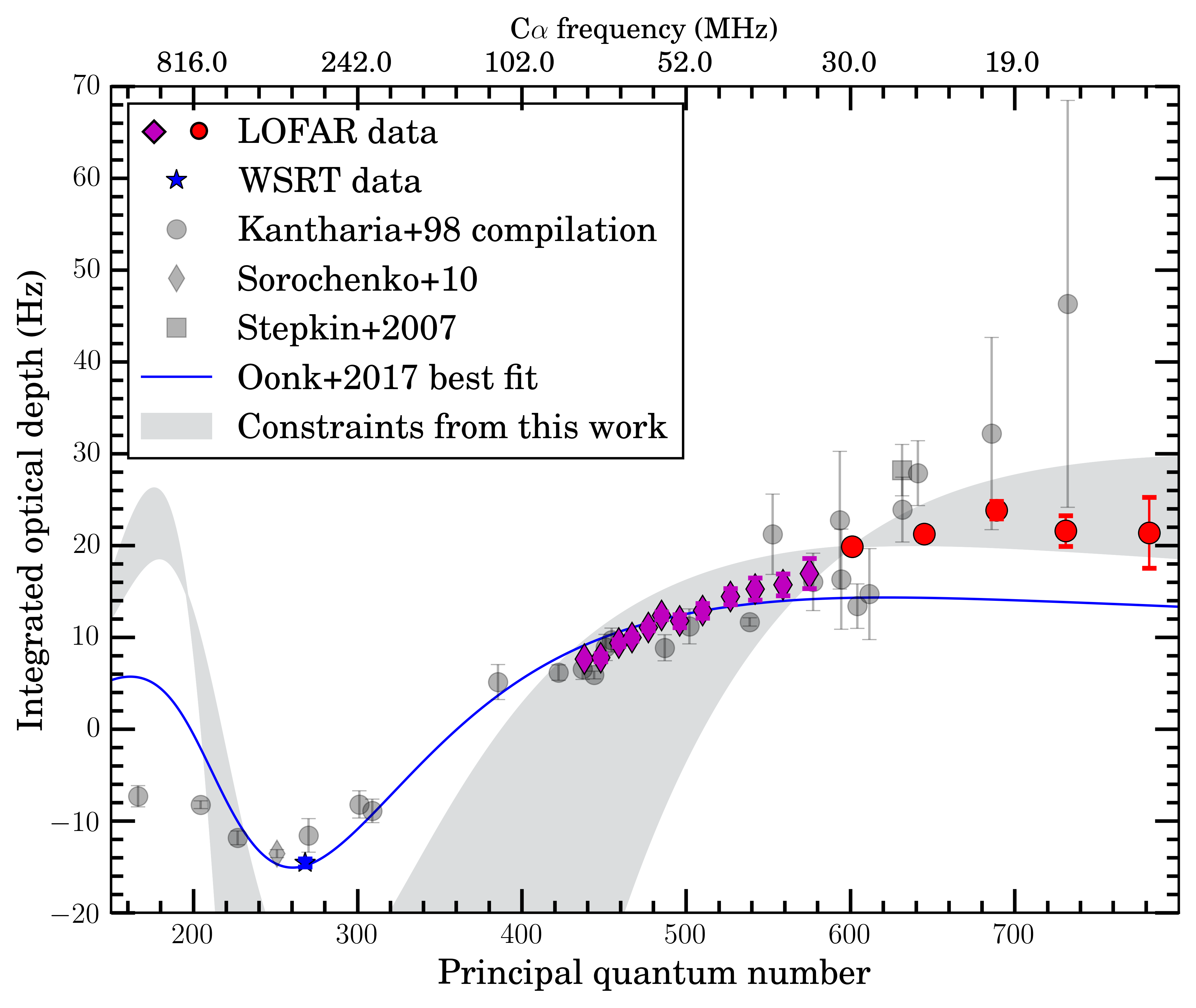}
 \end{center}
 \caption{Integrated optical depth as a function of principal quantum number for the sum of the Perseus arm components at $-47$ and $-38$~km~s$^{-1}$. 
 The LCASS data is shown with {\it red circles} for the data presented in this work and {\it purple diamonds} for the higher frequency data \citep{Oonk2017}. 
 The LCASS WSRT data is shown with a {\it blue star}. 
 Data from the literature is shown in {\it grey}.
 The best fit model from \citet{Oonk2017} for the sum of the $-47$ and $-38$~km~s$^{-1}$ components is shown with a {\it blue line}.
 The models which correspond to the physical constraints from this work are shown in a {\it gray shaded region}.
 In some cases the error bars are smaller than the symbols.}
 \label{fig:itau}
\end{figure}

The integrated optical depth as a function of $n$ is shown in Figure~\ref{fig:itau}.
In this Figure we also show the latest compilation of CRRL observations towards Cas~A presented in \citet{Kantharia1998}, the C$\alpha$ data points from \citet{Stepkin2007} and \citet{Sorochenko2010}, and the LCASS data \citep{Oonk2017}. 
The LOFAR points at $n>600$ show a fairly constant integrated optical depth as $n$ increases up to $800$. 
The observed shape of the change in integrated optical depth is similar to the one predicted by the models of CRRL emission \citep{Salgado2016a}. 
In these models a constant optical depth is reached faster than in previous ones \citep[e.g.,][]{Payne1994}.
A constant integrated optical depth is expected if the population of carbon atoms is close to LTE ($b_{n}\sim1$). 
When $b_{n}$ gets close to unity, $\beta_{n^{\prime}n}$ has an almost constant value. 
This is reflected in a constant integrated optical depth as a function of $n$. 
The models of \cite{Salgado2016a} show that this happens for $n>600$ in regions that satisfy $n_{e}>0.015\;\rm{cm}^{-3}(T_{e}/60\;\rm{K})^{-2}$.
Under these conditions, the highest $n$ levels are close to collisional equilibrium while the lower $n$ levels decay rapidly radiatively. 
As density or temperature increases, a larger fraction of the levels is close to collisional equilibrium and hence the total population of the collisionally dominated levels increases. 
This translates into an increment in the optical depth of the high $n$ transitions.

The LCASS data shows a similar trend when compared with previous results, but with a smaller scatter. 
LOFAR, with its large fractional bandwidth and high spectral resolution, is an ideal instrument for low frequency CRRL observations. 
Additionally, systematic differences arising from different calibration and data reduction strategies are minimised when using the same instrument. 
The trend in line width and integrated optical depths shows a smooth change with $n$. 
This suggests that there are no major systematic differences between the different LCASS LBA data sets. 
This is not the case for the literature data, for which a larger point to point scatter is observed.

Another advantage of LOFAR for CRRL observations is given by the frequency coverage of its subbands.
The frequency coverage of the LOFAR subbands results in sufficient velocity coverage to measure the Lorentzian wings of the line profiles for $n\lesssim800$. 
This was not the case for older receivers. 
As noted by \citet{Payne1994} some of the previous decametric observations could have underestimated the line profiles. 
This because in previous observations the velocity coverage of the spectrometers was comparable to the line width.
In this case, there are no line-free channels from which to determine the continuum.
This led \citet{Payne1994} to estimate how much the line profiles could have been underestimated and derive a correction based on this. 
However, such a correction depends on the model used for its derivation and the physical conditions assumed. 
Models which predict larger line widths will require larger correction factors for a fixed velocity coverage. 
Given that the gas conditions are not know a priori, this kind of correction can lead to erroneous determination of the gas properties.

\subsection{Carbon lines ratio}

A way of breaking the degeneracy between different gas properties of the observed CRRL properties is to compare the CRRLs with the $158$~$\mu$m [CII] line 
\citep{Natta1994,Salgado2016b}. Both lines are emitted by ionised carbon, but the line strengths have different temperature dependencies. This makes the 
CRRL/[CII] line ratio a good thermostat. 
Though, this also means that emission from both tracers can have different contributions from different phases of the ISM. 
The contribution from different phases of the ISM to the [CII] line intensity is still poorly constrained \citep[e.g.,][]{Pineda2013}. 
In the case of CRRLs this is still an open question \citep[see e.g.,][]{Anantharamaiah1994,Sorochenko1996}. 
Since the derived temperature for the CRRL gas is $85$~K \citep{Oonk2017}, which is close to the temperature required to excite the carbon 
$^{2}P_{3/2}$-$^{2}P_{1/2}$ core, here we will assume that both lines have equal contributions from different ISM phases.

To model the line intensity of the [CII] line we use the equations of \citet{Salgado2016b}. 
Given the gas conditions \citep{Oonk2017} and the atomic hydrogen column density \citep{Mebold1975} the [CII] line will be optically thick. 
We assume that the [CII] emission in the PACS spectra comes from the two velocity components at $-47$ and $-38$~km~s$^{-1}$. 
For the line widths we use $3.4$ and $6.8$~km~s$^{-1}$ respectively \citep{Oonk2017}. 
Additionally, we assume that the path length of the $-38$~km~s$^{-1}$ velocity component is half that of the $-47$~km~s$^{-1}$, as determined from higher frequency CRRL observations \citep{Oonk2017}. 
Under these assumptions we evaluate the intensity of the [CII] line in a $n_{e}$-$T_{e}$ grid which is the same as for the CRRLs.

Towards Cas~A, the ratio between the integrated optical depth of the C$(731)\alpha$ line and the [CII] line is $(-3.7\pm0.1)\times10^{5}$~Hz/(erg~s$^{-1}$~cm$^{-2}$~sr$^{-1}$).
The region of parameter space consistent with this ratio to within $3\sigma$ is shown in Figure~\ref{fig:pe} with red lines. 
The gas electron temperature is constrained to the range $T_{e}=70$--$100$~K for the lowest densities in our grid ($n_{e}<0.02$~cm$^{-3}$). 
For higher densities the allowed temperature range narrows considerably, decreasing to $60$--$70$~K for the highest density in our grid ($n_{e}=0.11$~cm$^{-3}$).
As the temperature increases the number of carbon atoms in the $^{2}P_{3/2}$ state is larger. 
This translates into an increase of the [CII] line intensity. 
In practice this means that for a smaller CRRL/[CII] ratio the allowed temperature would be higher. 

We do recognise that the [CII] $158$~$\mu$m intensity used only refers to $\sim20\%$ of the Cas~A supernova remnant and is unresolved in velocity. 
As shown by \citet{Gerin2015}, observations of [CII] unresolved in velocity can hide absorption features. 
This has the effect of lowering the observed [CII] intensity when absorption and emission features are convolved with the spectrometer response.
Further high spectral resolution observations with SOFIA are necessary to determine this ratio well.

\subsection{CRRL ratio}

A way in which the emission measure dependence is eliminated is by considering the ratio between the integrated optical depth of CRRLs. 
The ratio between CRRLs has a weaker dependence on calibration uncertainties.
The measured C$\alpha(731)$/C$\alpha(601)$ integrated optical depth ratio has a value of $1.12\pm0.07$.
The region of parameter space which produces a line ratio consistent with the observations to within $3\sigma$ is shown in Figure~\ref{fig:pe} as the hatched 
region between yellow lines. 
Even if the range of pressures allowed by the $3\sigma$ values is large, it does pose a strict lower limit on the electron pressure of $P_{e}\geq0.9$~K~cm$^{-3}$, for $T_{r,100}=800$~K and $T_{e}\geq20$~K.
For higher values of $T_{r,100}$ the lower limit will be larger.
We can turn this in to a lower limit on the gas pressure by adopting an electron fraction. 
If we use an electron fraction equal to the gas phase carbon abundance \citep[$1.5\times10^{-4}$,][]{Cardelli1996,Sofia1997}, i.e. all the electrons come from ionized carbon, this is $P\geq6\times10^{3}$~cm$^{-3}$~K.
This gas pressure is comparable to that measured using UV absorption lines of neutral carbon \citep{Jenkins2001}.

\subsection{Piecing it all together}
\label{ssec:pat}

\begin{figure*}
 \begin{center}
  \includegraphics[width=0.99\textwidth]{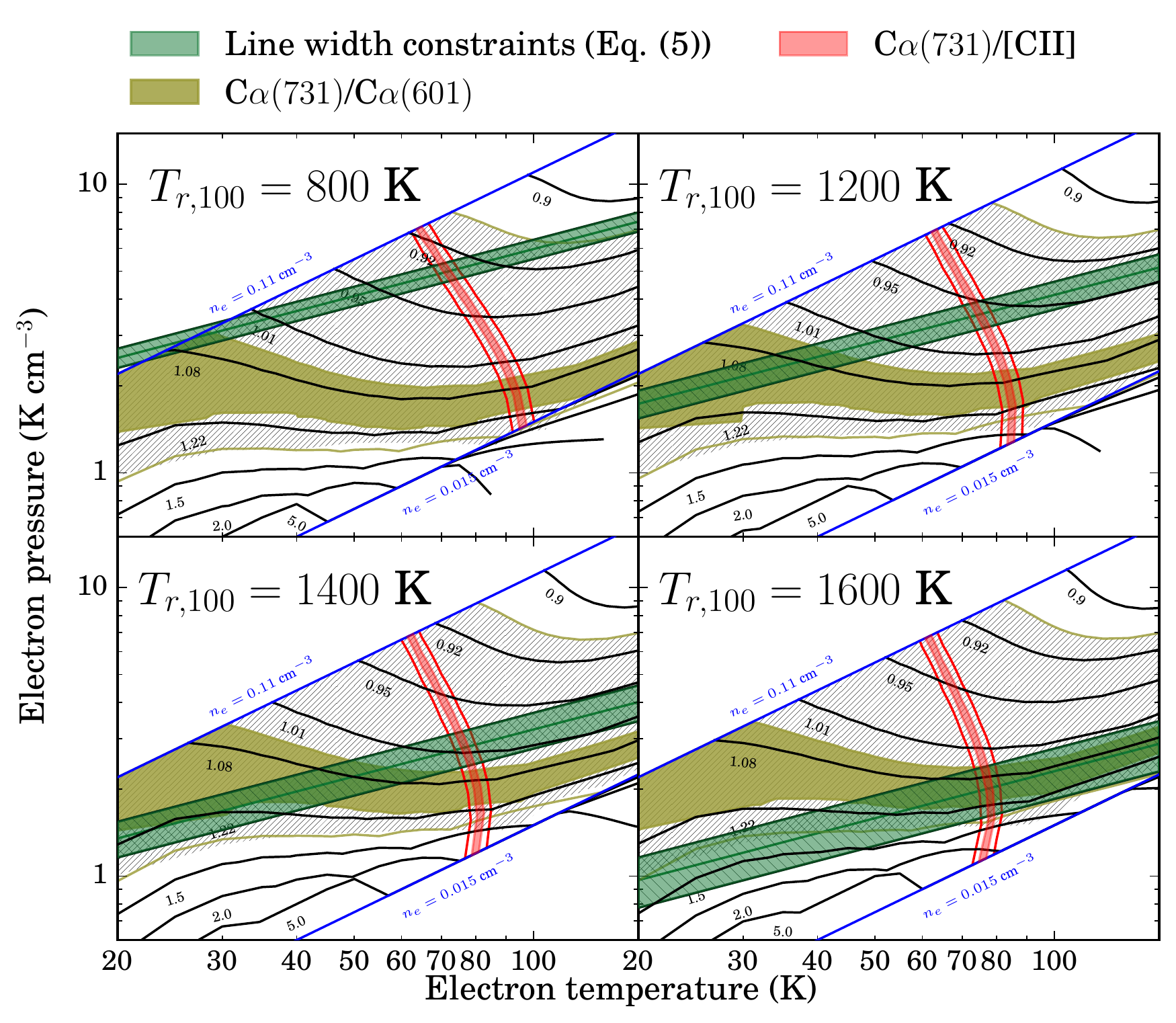}
 \end{center}
 \caption{Pressure temperature diagrams for CRRLs below $33$~MHz and the $158$~$\mu$m [CII] line. 
 These can be used to constrain the gas properties given measurements of decametric CRRLs and [CII]. 
 The {\it black lines} show the pressure for constant values of the C$\alpha(731)$/C$\alpha(601)$ ratio. 
 The {\it red region} shows the allowed temperature range derived from the C$\alpha(731)$/[CII] ratio with a $1\sigma$ error.
 The {\it red lines} show the allowed temperature range when considering a $3\sigma$ error bar.
 The {\it green region} shows the allowed range of pressures given Eq.~(\ref{eq:lwrel}) for $T_{r,100}\pm100$~K. 
 The {\it yellow region} shows the measured value of the C$\alpha(731)$/C$\alpha(601)$ ratio with a $1\sigma$ error, while the {\it densely hatched region} 
shows the same value with a $3\sigma$ error.
 The difference between the $1\sigma$ and $3\sigma$ curves reflects the importance of having high signal-to-noise detections. 
 The {\it blue lines} show constant electron density curves for $n_{e}=0.015$ and $0.11$~cm$^{-3}$.
 For values of $T_{r,100}>1650$~K, the green region goes outside the allowed range of parameters.}
 \label{fig:pe}
\end{figure*}

Using the measured temperature from the CRRL--[CII] ratio, and the constraints from the C$\alpha(731)$/C$\alpha(601)$ ratio and line width we can break the degeneracy between $T_{e}$, $n_{e}$ and $T_{r,100}$.
We can combine these constraints since the line width traces the most prominent velocity component at $-47$~km~s$^{-1}$ and we assume that the $-38$~km~s$^{-1}$ velocity component traces gas with similar physical properties.
An example of this is shown in Figure~\ref{fig:pe}.
In this Figure we can see how the line width relation (Eq.~(\ref{eq:lwrel})) and the pressure derived from the C$\alpha(731)$/C$\alpha(601)$ ratio intersect only at the correct value of $T_{r,100}$, when we adopt the $1\sigma$ value for the C$\alpha(731)$/C$\alpha(601)$ ratio. 
Then, from the allowed range of pressure and temperature the density can be determined. 
However, if we consider the $3\sigma$ range for the C$\alpha(731)$/C$\alpha(601)$ ratio, then we are in a situation were $T_{r,100}$ and the pressure are degenerate.
This shows the importance of having high signal-to-noise data to determine the gas conditions.

Using the ranges allowed by our data we note that a value of $T_{r,100}$ lies between $1500$ and $1650$~K, $T_{e}$ between $68$--$98$~K and $P_{e}$ between $1.5$--$2.9$~K~cm$^{-3}$. 
The density range allowed by the intersection of these constraints is $0.02$--$0.035$~cm$^{-3}$. 
These results are in line with those published by \citet{Oonk2017}, but here we have made use only of CRRLs below $33$~MHz and the $158$~$\mu$m [CII] line.

In Figure~\ref{fig:itau} we also show how the derived range of physical conditions translate into integrated optical depth.
We can see how the LOFAR data falls within the range of allowed physical conditions.
However, the integrated optical depth is underestimated for the WSRT data point.
When performing a spatial average, such as that done inside the synthesized beam of the observations presented here, the observed line optical depth will be a weighted average of the line optical depth. 
The weight is proportional to the brightness temperature of the background source.
If the background source shows spectral index variations on scales smaller than the synthesized beam, then the line profiles will sample different spatial structures with frequency.
The continuum from Cas~A shows variation in its optical depth on scales of arcseconds, with a flatter spectral index in its inner $50\arcsec$ \citep{DeLaney2014}.
The spatial structure of the gas at $-47$~km~s$^{-1}$ shows variations on arcminute scales with the gas concentrated towards the south and west of Cas~A \citep{Anantharamaiah1994,Asgekar2013}.
The integrated optical depth towards the western hotspot of Cas~A can be a factor of two larger than that extracted over the whole face of Cas~A \citep{Asgekar2013}.
This implies that at frequencies below $30$~MHz the line profiles will be weighted towards the gas in the south and west of Cas~A.
Similarly, gas outside the supernova remnant could be sampled by the lower frequency observations.
However, we estimate that the surface brightness of Cas~A relative to the surrounding Milky Way at these frequencies is larger by a factor of $\gtrsim10$. 
Then, the effect of gas outside the face of Cas~A should be small unless its integrated optical depth is ten times larger than that of the gas which covers the face of Cas~A.
Another reason for underestimating the WSRT integrated optical depth is related to the change of physical conditions traced by CRRLs at different frequencies.
This can be caused because the CRRL optical depth acts as a density filter \citep[e.g.][]{Mohan2005}, or because some components can be absorbed by free electrons at lower frequencies \citep[e.g.][]{Pankonin1980}.
CRRLs act as a density filter because gas with different physical conditions will emit/absorb preferentially at different frequencies.
A model with a single slab of gas does not take into account this.
However, the good fit of \citet{Oonk2017} to the integrated optical depth of the $-47$~km~s$^{-1}$ velocity component using a single slab of gas suggests that along this line of sight low frequency ($\nu\lesssim1$~GHz) CRRLs trace gas with a narrow range of physical conditions.
The last possibility we consider is that the physical conditions of the $-47$ and $-38$~km~s$^{-1}$ gas are not the same.
The work of \citet{Oonk2017} at higher frequencies shows that the $-38$~km~s$^{-1}$ velocity component is less well fit than the $-47$~km~s$^{-1}$ one.
However, with the current data we do not find a significant difference between the two components.
Data at higher frequencies, e.g., obtained with the LOFAR HBA ($120$--$240$~MHz), would help narrow down the physical conditions of the $-38$~km~s$^{-1}$ velocity component.

\subsection{The Perseus arm Cas~A distance}

In the previous subsections we have obtained a range of allowed values for $T_{r,100}$.
Even if we can not accurately determine the contribution from Cas~A to the radiation field it is interesting to use the constraints on $T_{r,100}$ to try and 
determine the distance between the Perseus arm gas and the background source.
Figure~\ref{fig:pe} shows that $T_{r,100}$ can take values between $1500$ and $1650$~K if we consider the $1\sigma$ ranges.
The brightness of the source at $100$~MHz extrapolated from the \citet{Vinyaikin2014} model is $1.8\times10^{7}$~K.
If we adopt a contribution from the Milky Way to $T_{r,100}$ of $800$~K and a radius of $3$~pc for Cas~A, then the distance between the Perseus arm gas at 
$-47$~km~s$^{-1}$ and Cas~A can be between $262$ and $220$~pc.
This distance would increase if the contribution from the Milky Way to $T_{r,100}$ is larger.

We can compare this with estimates of the distance to the Perseus arm gas based on the Galactic rotation curve. 
We use a flat rotation curve outside the solar circle ($R_{\odot}=8.5$~kpc) with a constant velocity of $220$~km~s$^{-1}$ \citep{Dickey2009}. 
For the $-47$~km~s$^{-1}$ component, the distance of the Perseus arm gas to Earth is $4.5$~kpc. 
This distance is larger than the Earth-Cas~A distance \citep[$3.4$~kpc][]{Reed1995}, but this cannot be the case as the lines appear in absorption. 
Parallax measurements of star forming regions in the Perseus arm show that kinematic distances are in disagreement with parallax determined distances \citep[e.g.,][]{Xu2006,Choi2014}. 
Based on these measurements the distance from Earth to the gas in the Perseus arm towards Cas~A is $\approx3.3$~kpc.

\section{Conclusions}
\label{sec:conclusions}

We have observed Cas~A in the $10$--$33$~MHz LBA band of LOFAR. 
Our results show that, while observations are hampered by ionospheric effects and RFI noise, C$\alpha$ lines can be readily detected over this frequency range up to principal quantum numbers in excess of $\sim800$ and C$\beta$ and C$\gamma$ lines to even higher quantum numbers. 
At the lowest frequencies, analysis of this data is hindered because of severe line broadening, which leaves a dearth of line free channels. 
This limits the effective analysis of C$\alpha$ lines to $n\lesssim800$.
Nevertheless, this still gives access to hundreds of transitions. 

As line frequencies are accurately known, stacking procedures can be used to increase the signal to noise and very weak lines ($\sim10^{-3}$ in peak optical depth) can still be reliably detected ($S/N>10$) after stacking. 
These lines are caused by absorption associated with the Perseus spiral arm along the line of sight towards this supernova remnant. 
The observed lines show a pronounced broadening with principal quantum number while the peak optical depth decreases. 
The integrated optical depth stays quite constant over the full n-range ($600$--$800$). This contrasts with previous observations which 
suggested a continuous increase in integrated optical depth with principal quantum number.
The line broadening at high $n$ reflects the importance of radiation and/or pressure broadening and provides insight into the physical conditions of the emitting gas.
We find that the change in line width with frequency can be described by a model in which the radiation field is due to Galactic synchrotron emission, without the necessity to invoke additional contributions.
However, deviations from a power law in the radiation broadening term are difficult to measure given the biases in the line recovery process at these low frequencies.
Higher signal-to-noise observations would help to determine if the apparent flattening in the line width at the highest principal quantum numbers is significant.

The high-$n$ lines presented in this work are close to collisional equilibrium and a constant integrated optical depth is expected.
As they are close to LTE, the ratio of the CRRLs is not very sensitive to the physical conditions but they do put a firm constraint on the minimum gas electron pressure of $P_{e}\geq0.9$~K~cm$^{-3}$, which corresponds to $6\times10^{3}$~K~cm$^{-3}$ for an electron fraction of $1.5\times10^{-4}$.

By using the CRRL-[CII] line ratio as a thermostat, the ratio between CRRLs as a barometer and the line widths we are able to obtain physical conditions for the gas.
Based on the data in the range $10$--$33$~MHz and its $1\sigma$ errors we derive $T_{e}=60$--$98$~K, $T_{r,100}=1500$--$1650$~K and $n_{e}=0.02$--$0.035$~cm$^{-3}$.
This highlights the power of combining low frequency CRRLs with the $158$~$\mu$m [CII] line.
However, we caution that these results are uncertain due to the blend of at least two velocity components along this line of sight.
Adding to this uncertainty is the unknown relation between [CII] and low frequency CRRLs.
Also, due to the limited spatial coverage and spectral resolution of the Herschel PACS observations, the [CII] intensity is not very well known. 
High spectral and spatial resolution observations with SOFIA would greatly help to determine the [CII] intensity over the face of Cas~A. 

Our study demonstrates that the lowest frequency window available with LOFAR can provide important insights into the physical conditions of the gas traced by CRRLs.

\section*{Acknowledgements}
{We would like to thank the anonymous referee for useful comments which helped improve the quality and readability of this manuscript.
P.~S., J.~B.~R.~O., A.~G.~G.~M.~T, H.~J.~A.~R. and K.~E. acknowledge financial support from the Dutch Science Organisation (NWO) through TOP grant 614.001.351.
LOFAR, designed and constructed by ASTRON, has facilities in several countries, that are owned by various parties (each with their own funding sources), and 
that are collectively operated by the International LOFAR Telescope (ILT) foundation under a joint scientific policy. {\it Herschel} is an ESA space observatory 
with science instruments provided by European-led Principal Investigator consortia and with important participation from NASA. We gratefully acknowledge that 
LCASS is carried out using Directors discretionary time under project DDT001.
R.~J.~W. is supported by a Clay Fellowship awarded by the Harvard-Smithsonian Center for Astrophysics. A.~G.~G.~M.~T acknowledges support through the Spinoza 
premie of the NWO.}

{\it Facilities:} LOFAR, Herschel.

\bibliographystyle{mnras}
\bibliography{ref_rrl.bib}

\appendix

\section{Simulated line profiles}
\label{app:sim}

\begin{figure*}
 \begin{center}
  \includegraphics[width=0.99\textwidth]{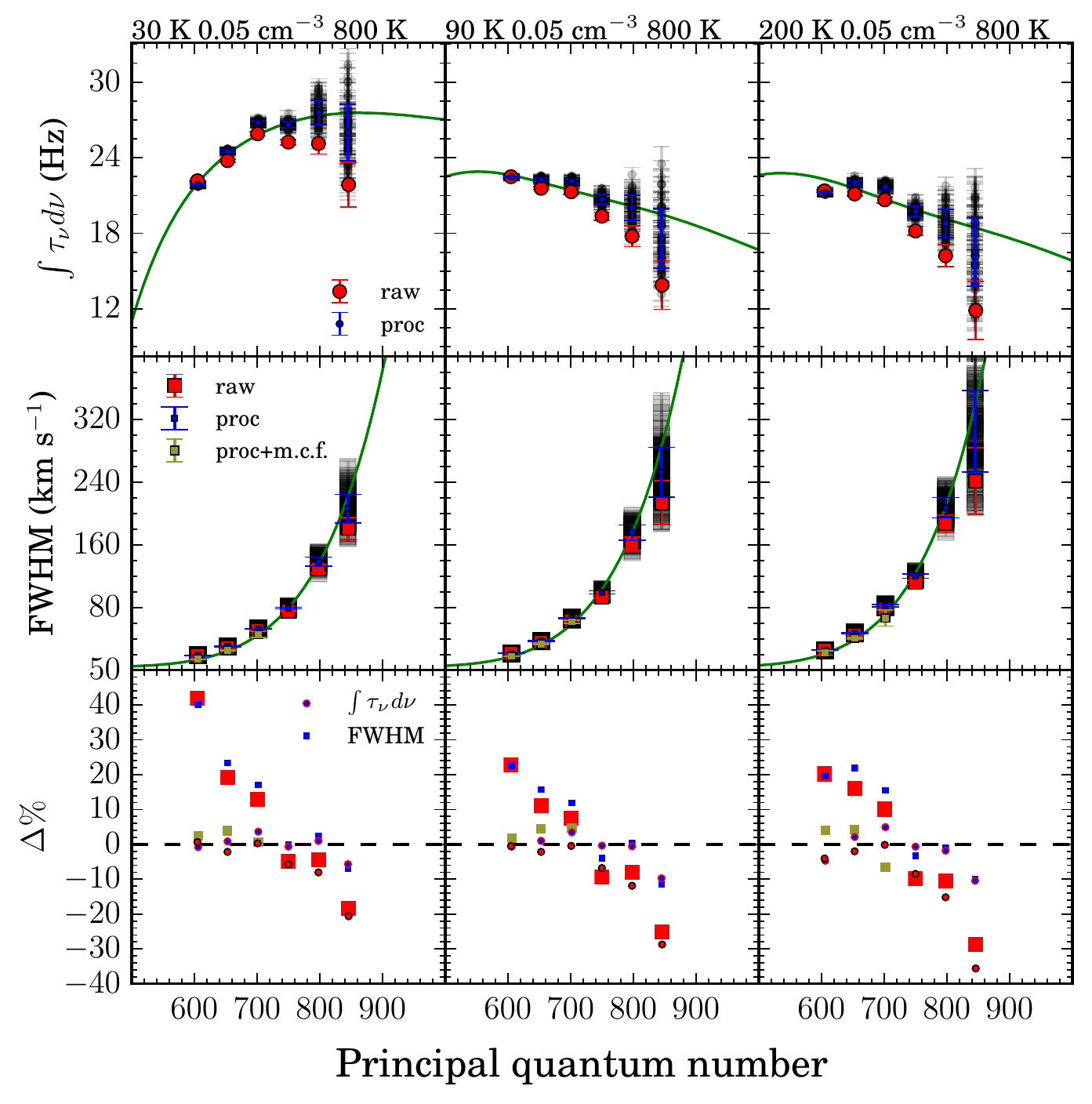}
 \end{center}
 \caption{Comparison between the input line properties and the measured line properties of synthetic spectra. 
 The {\it top} row shows the integrated optical depth as a function of principal quantum number. 
 The {\it grey points} show the results of individual simulations after continuum subtraction, stacking and baseline removal, the {\it blue points} show the mean of the grey points and the {\it red points} show the mean of the stacks with no baseline removal. 
 The input line property is shown with a {\it green line}.  
 The {\it middle} row shows the line FWHM. 
 The symbols have the same meaning as in the top row, but the points are now {\it squares}. 
 The {\it yellow squares} show the line width of the dominant component when multiple velocity components are fitted (m.c.f.) to the line profiles.
 The {\it bottom} panel shows the difference between the input value and the measured one. 
 The comparison with processed data is shown by the {\it blue symbols}, while for the unprocessed data we use {\it red symbols}.
 The {\it yellow squares} show the difference between the input model for the dominant component and the measured line width for the dominant component after processing.}
 \label{fig:sims}
\end{figure*}

We generate synthetic spectra using the same number of channels and bandwidth as the LOFAR observations. 
To each synthesised spectrum we add Gaussian noise that mimics the noise properties of the LOFAR data. 
Ripples that simulate the effect of the bandpass response of LOFAR \citep{Romein2008} are also added. 
Then we add C$\alpha$, C$\beta$, C$\gamma$ and C$\delta$ lines to the spectra using Voigt profiles and generating the line properties based on the \citet{Salgado2016a} models. 
For each transition and for each velocity component we add one Voigt profile. The code used can be found in \url{https://github.com/astrofle/CRRLpy} \citep{crrlpy}. 
The velocity components are centred at $-47$, $-38$ and $0$~km~s$^{-1}$. 
The synthesised spectra with the line profiles are then converted to antenna temperatures by inverting Eq.~\ref{eq:tau} to add the continuum back. 
In all cases we used a cloud covering factor of unity and a Doppler line width of $3.4$~km~s$^{-1}$. 
We use $3$ different temperatures for the line properties, $30$~K, $90$~K and $200$~K. 
The emission measure is adjusted to match the observed integrated optical depths. 
For the $30$~K model the emission measure of the $-47$, $-38$ and $0$~km~s$^{-1}$ components is $0.03$, $0.02$ and $0.0045$~pc~cm$^{-6}$ respectively. 
For the $90$~K model these are $0.08$, $0.04$ and $0.009$~pc~cm$^{-6}$, for the $200$~K model $0.25$, $0.1$ and $0.05$~pc~cm$^{-6}$. 
The synthetic spectra are processed in the same way as the LOFAR spectra. 
First we remove the continuum by estimating it from line free channels and then the stacking and baseline removal process is performed. 
This was repeated $60$ times. 
Each time the noise in the synthetic spectra was recomputed.
A comparison between the input models and the measured values from the stacks is shown in Figure~\ref{fig:sims}. 
Here we also compare with stacks that were not processed, i.e., C$\alpha$, C$\beta$, C$\gamma$ and C$\delta$ lines are present in the spectrum while stacking and no bandpass correction was applied. 

In the top panel of Figure~\ref{fig:sims} we compare the integrated optical depth. 
This comparison shows that before processing the integrated optical depth is underestimated for ${n>710}$. 
We have investigated the cause of this bias and have concluded that it arises because the number of lines per frequency unit increases with $n$. 
At ${n\approx710}$ the Lorentzian wings of adjacent CRRLs begin to overlap and correctly estimating the continuum becomes difficult. 
To test this we generated the same line profiles with a constant noise level of $10^{-5}$, in optical depth units, and added only C$\alpha$ lines. 
No processing other than continuum removal and stacking was applied. 
The results are very similar, indicating that the underestimation is due to the presence of C$\alpha$ lines in the spectra. 
After processing, the measured values are closer to those of the input model. 
The difference between the integrated optical depth of the input model and the measured value after processing can be up to $16\%$, in the worst case.

The middle row of Figure~\ref{fig:sims} shows a comparison between the measured line width and the line width of the dominant component, at $-47$~km~s$^{-1}$.
The measured line full width at half maximum (FWHM) is overestimated for ${n<710}$. This is partially because for ${n<710}$ the Orion component is still not 
completely blended with the Perseus components, so that fitting a single Voigt profile will result in a larger line FWHM. In the middle and bottom panels of 
Figure~\ref{fig:sims} we also show the measured value when fitting up to $3$ Voigt profiles. The number of profiles used is such that the fit residuals are 
minimised. For ${n>710}$ the profiles are blended enough that fitting a single Voigt profile is enough. Also, the results of the simulations show that after 
processing the line width is underestimated by no more than $\sim16\%$ for ${n>710}$. These results also show that even if the different velocity components 
are blended, the measured line width will resemble the line width of the most prominent velocity component.

We also analysed the results for C$\beta$ and C$\gamma$ lines. After our processing, Sect.~\ref{ssec:stack}, the integrated optical depth and line width of 
C$\beta$ lines can be recovered to within $3\%$ and $10\%$ respectively for ${n<800}$. For C$\gamma$ lines the measured values can be recovered to within 
$50\%$ and $30\%$ for the integrated optical depth and line width respectively. In both cases using a single Voigt profile when fitting provides the best 
results. Based on these results we choose to consider only the C$\beta(761)$ detection in the analysis.

These simulations do not include an effect which affects the data. This is contamination and loss of channels due to RFI. In some cases strong RFI is present on 
top of the lines, and the affected channels have to be flagged resulting in an incomplete line profile. This is partially alleviated by the fact that the stack 
contains multiple lines which suffer from RFI differently. Even though we have not investigated the effect of RFI in the final stacks, we believe this 
will have a minor effect on the derived line properties.\\

\bsp    
\label{lastpage}
\end{document}